\documentclass[parskip=full, british]{scrartcl}
\usepackage[utf8]{inputenc}
\usepackage{float}
\usepackage{amsmath}
\usepackage{amsthm}
\usepackage{amssymb}
\usepackage{algorithm}
\usepackage{algpseudocode}
\usepackage{geometry}
\usepackage{graphicx}
\usepackage{booktabs}
\usepackage{url}
\usepackage[round]{natbib}
\usepackage[nomarkers,notablist,nofiglist,tablesfirst]{endfloat}
\setcitestyle{citesep={,},aysep={}}

\theoremstyle{plain}
\newtheorem{theorem}{Theorem}[section]
\newtheorem{lemma}[theorem]{Lemma}
\newtheorem{proposition}[theorem]{Proposition}
\theoremstyle{definition}

\theoremstyle{remark}
\newtheorem{remark}[theorem]{Remark}

\usepackage{babel}
\date{}

\begin{document}

\title{European derivatives pricing via constrained Gaussian process
regressions}
\author{Arnaud Servais\thanks{Corresponding author. E-mail: arnaud.servais@uclouvain.be}\\
{\normalsize\textit{Institute of Statistics, Biostatistics and Actuarial Sciences (ISBA), LIDAM,}}\\
{\normalsize\textit{UCLouvain, Louvain-la-Neuve, Belgium}}\\[6pt]
Donatien Hainaut\thanks{E-mail: donatien.hainaut@uclouvain.be}\\
{\normalsize\textit{Institute of Statistics, Biostatistics and Actuarial Sciences (ISBA), LIDAM,}}\\
{\normalsize\textit{UCLouvain, Louvain-la-Neuve, Belgium}}}
\maketitle

\begin{abstract}
This article develops a Gaussian process method for pricing European derivatives when the underlyings, and the risk factors driving them, follow a multivariate jump-diffusion.  A Gaussian process prior on the pricing function is
conditioned on the Feynman--Kac partial integro-differential equation at a set
of collocation points. The price is returned as a closed-form and differentiable function, allowing the Greeks to be obtained without simulation or finite differences. The jump terms are handled iteratively: the price is computed as the limit of a sequence of boundary value problems. Our main theoretical contribution is a convergence result for this
iteration: using Weissinger's fixed-point theorem in the supremum norm, we show
that it converges from any starting point, and provide a worst-case number of iterations for a given precision. The method is illustrated with derivatives of two underlyings under the Heston model with Kou double-exponential jumps, a five-dimensional problem, and benchmarked
against Monte Carlo simulation.

\\

\textit{Word count}: 6097
\end{abstract}

\clearpage
\section{Introduction}
\label{sec:intro}

This paper introduces an iterative algorithm based on constrained Gaussian
process regression for pricing financial derivatives when the underlying
dynamics include jumps and stochastic risk processes influencing the underlying (e.g. stochastic volatility). Continuous diffusion models
yield tractable pricing equations, but empirical returns exhibit heavy tails,
excess kurtosis and pronounced volatility smiles that a pure diffusion cannot
reproduce \citep{merton1976option,kou2002jump,cont2003financial}. Adding a
jump component helps restore agreement with market observations, at the cost of a
pricing equation that is no longer purely differential. By the infinitesimal
generator and the Feynman--Kac theorem, the pricing function
$V(t,\boldsymbol y_{t})$ of a derivative written on a state vector
$\boldsymbol x=(t,\boldsymbol y_{t})\in\mathcal X\subset\mathbb R^{p}$
satisfies a partial integro-differential equation
\begin{equation}\label{eq:intro_pide}
  \mathcal L_{\boldsymbol x}V(\boldsymbol x)+\mathcal J V(\boldsymbol x)=0 ,
\end{equation}
subject to the boundary condition $V(\boldsymbol x)=h(\boldsymbol x)$ carrying
the terminal payoff of the instrument. Here $\mathcal L_{\boldsymbol x}$ is a
local second order differential operator describing the continuous part of the
dynamics and $\mathcal J$ is a non-local integral operator describing the jump
transitions.

Gaussian process regression is a Bayesian nonparametric method long established
in statistics and machine learning \citep{rasmussen2006gaussian}. Its use as a solver rather than as a regressor rests
on a structural property: a linear operator maps a Gaussian process to a jointly
Gaussian process, so a process can be conditioned on the value of a differential
operator applied to it. This observation is due to \citet{graepel2003solving}
and was developed for stochastic partial differential equations by
\citet{sarkka2011linear} and into a general theory of probabilistic numerics by
\citet{owhadi2015bayesian}, and more recently
into a framework for solving and learning nonlinear partial differential
equations with kernels \citep{chen2021solving}. The same idea underlies the
physics-informed machine learning literature \citep{raissi2019physics}, of which
\citet{gulian2019machine} give a Gaussian process formulation for fractional
operators. In quantitative finance, Gaussian processes and neural networks have mainly been used as
surrogates rather than as solvers: \citet{de2018machine} fit prices and Greeks produced by an
exogenous pricer, \citet{goudenege2020machine} use Gaussian process regression for American options
in high dimension, and \citet{crepey2020gaussian} price
derivative portfolios for counterparty risk. Chapters 4 and 5 of
\citet{ludkovski2025gaussian} survey this body of work. What
distinguishes the constrained formulation used here is that no exogenous price
is required at any stage: the pricing equation itself supplies the training
information.

Monte Carlo
methods \citep{glasserman2004monte} approximate the price at
one state at a time and produce sensitivities with sampling noise. The relevant
alternatives are deterministic solvers of \eqref{eq:intro_pide}, and their
common difficulty is precisely the non-local term, which destroys the banded
structure that makes implicit time stepping cheap. Three strategies have
emerged. The first evaluates the integral explicitly and the differential part
implicitly, giving the implicit-explicit schemes of
\citet{andersen2000jump} and \citet{cont2005finite}, with convergence and
stability analysed in the viscosity framework by the latter.
Method-of-lines solvers also handle two assets with stochastic
volatility and jumps \citep{garces2021numerical}. The second treats the integral by a fixed-point iteration nested inside each
implicit time step, with fast transform evaluations of the integral
\citep{toivanen2008numerical,salmi2014imex}; there the iteration contracts
because the effective intensity over one time step is small. The third exploits an explicit
characteristic function, as in the Fourier cosine expansion of
\citet{fang2009novel}, which is extremely fast, but in low dimension. A fourth line
of work, closest in spirit to what follows, iterates over the number of jumps
rather than over time steps: \citet{bayraktar2009pricing} approximate the
American price under jump diffusion by a sequence of optimal stopping problems
for the jump-free diffusion and show that the sequence converges uniformly and
exponentially fast. The method proposed here requires
only pointwise evaluation of the jump measure, and returns an analytic function
of the state together with a posterior variance.

Extending constrained Gaussian processes to jump-diffusion dynamics is not
immediate, because the non-local operator $\mathcal J$ rarely admits a closed-form
action on the covariance kernel. Embedding it in the prior would require a
quadrature for each of the entries of the Gram matrix, and a
fresh factorisation of that matrix whenever the jump parameters change. We
therefore keep $\mathcal J$ out of the prior altogether and treat the jump term as a
lagged source. The jump expectation is collected in the operator
$\mathcal J(V)(\boldsymbol x)=\sum_{v}\lambda_{v}\mathbb E[V(t,\boldsymbol y+M^{(v)}\xi^{(v)})]$,
and at iteration $n\ge1$ the price $V^{(n)}$ solves
\begin{equation}\label{eq:intro_iter}
  \mathcal L_{\boldsymbol x}V^{(n)}(\boldsymbol x)=z^{(n-1)}(\boldsymbol x),
  \qquad z^{(n)}:=-\mathcal J\bigl(V^{(n)}\bigr),
\end{equation}
where $z^{(n)}$ is the jump integral evaluated on the previous surface. Each
line of \eqref{eq:intro_iter} is a purely local boundary value problem, solved
in closed form by the same constrained regression. Because the source enters
only the right-hand side, the Gram matrix is assembled and factorised once, and
the per-iteration cost of the jump term becomes linear in the entries of the Gram matrix, rather than quadratic.

Our contributions are the following. We extend constrained Gaussian process
regression from parabolic partial differential equations to the partial
integro-differential equations arising under multivariate jump diffusion,
through a splitting that keeps the jump operator entirely out of the Gram matrix, which considerably improves the numerical tractability. We then prove, using the fixed-point
theorem of \citet{weissinger1952theorie}, that the resulting fixed-point iteration converges in the
supremum norm, from any starting guess and without any restriction on the
maturity or on the jump intensity. We also give a worst case number of iterations required for a given precision. Finally,
we derive closed-form expressions for the Gaussian and Mat\'ern $5/2$ kernel families,
and validate the whole construction on the \citet{heston1993closed} model with \citet{kou2002jump} jumps against
Monte Carlo simulation.

The remainder of the paper is organised as follows.
Section~\ref{sec:model} introduces the multivariate jump-diffusion model and
Section~\ref{sec:fk} the corresponding Feynman--Kac partial integro-differential
equation, which Section~\ref{sec:bvp} restates as a boundary value problem on a
bounded domain. Section~\ref{sec:cgpr} presents the constrained Gaussian process
regression and the splitting of the jump term, and Section~\ref{sec:fixedpoint}
proves the convergence of the resulting fixed-point iteration, derives a
stopping criterion and describes its implementation. Section~\ref{sec:kernels}
derives the kernel images for radial kernels and specialises them to the
Mat\'ern~$5/2$ and Gaussian kernels, Section~\ref{sec:hyper} discusses the
selection of the kernels parameters, and Section~\ref{sec:greeks} gives the Greeks
in closed form. Section~\ref{sec:numerics} applies the method to a call, a put
and a double knock-out put on a basket of two assets under the Heston--Kou model.

\section{Jump-diffusion model}
\label{sec:model}

We consider a financial market driven by a
$(p-1)$-dimensional vector of risk
processes $\boldsymbol{y}_t$, defined on a complete probability space
endowed with a filtration $\mathcal{F} = (\mathcal{F}_t)_{t\geq0}$. We assume
these risk processes follow the stochastic differential equation
\begin{equation}\label{eq:JDmodel}
	d\boldsymbol{y}_t = \boldsymbol{\mu}(t, \boldsymbol{y}_t) dt
	+ \Sigma(t, \boldsymbol{y}_t) d\boldsymbol{W}_t
	+ M(t, \boldsymbol{y}_t)d\boldsymbol{L}_t
\end{equation}
where $\boldsymbol{\mu}(t, \boldsymbol{y}_t)$ is a $(p-1)$-vector,
$\Sigma(t, \boldsymbol{y}_t)$ and $M(t, \boldsymbol{y}_t)$ are
$(p-1)\times(p-1)$ matrices, $\boldsymbol{W}_t$ is a $(p-1)$-vector of
independent Brownian motions and $\boldsymbol{L}_t$ is a $(p-1)$-vector of
independent jump processes. Even though $\boldsymbol{W}_t$ and
$\boldsymbol{L}_t$ contain independent components, dependence between the
components of $\boldsymbol{y}_t$ is generated through the matrices
$\Sigma(t, \boldsymbol{y}_t)$ and $M(t, \boldsymbol{y}_t)$.

We model $\boldsymbol{L}_t$ as a vector of compound Poisson processes whose $v$-th component, for $v \in \{1, \dots, p-1\}$, is defined as
\begin{equation*}
  L^{(v)}_t = \sum_{j=1}^{N_t^{(v)}} \xi_j^{(v)},
\end{equation*}
where $(N_t^{(v)})_{t \geq 0}$ is a Poisson process with intensity $\lambda_v \in (0, \infty)$, and $(\xi_j^{(v)})_{j\ge1}$ is an independent sequence of i.i.d.\ jump sizes distributed as a generic random variable $\xi^{(v)}$ with finite moments of all orders.

We write $\Lambda:=\sum_{v=1}^{p-1}\lambda_v$ for the total jump
intensity of the system, a quantity that will govern the convergence rate of our scheme.

We impose the following integrability conditions on the coefficients:
\begin{alignat*}{2}
    & \mathbb{E} \left( \int_0^t \|\boldsymbol{\mu}(s, \boldsymbol{y}_s)\| \, ds \right) &&< \infty, \quad \forall t < \infty, \\
    & \mathbb{E} \left( \int_0^t \|\Sigma(s, \boldsymbol{y}_s)\Sigma(s, \boldsymbol{y}_s)^\top\|\, ds\right) &&< \infty, \quad \forall t < \infty, \\
    & \mathbb{E} \left( \int_0^t\|M(s, \boldsymbol{y}_s)M(s, \boldsymbol{y}_s)^\top\|\, ds \right) &&< \infty, \quad \forall t < \infty.
\end{alignat*}
As this paper aims to price and hedge derivatives, we assume an
arbitrage-free market and work under a risk-neutral measure $\mathbb{Q}$.
The coefficients $\boldsymbol{\mu}, \Sigma, M$ and the processes
$\boldsymbol{W}_t, \boldsymbol{L}_t$ are assumed to be specified so that the
discounted prices of the tradable securities in $\boldsymbol{y}_t$ are
$(\mathbb{Q}, \mathcal{F})$-martingales.

\section{Feynman--Kac equation}
\label{sec:fk}

We consider a European derivative whose price at any time $t\leq T$ is a
function of the current state, $V(t,\boldsymbol{y}_{t})$. Since the
discounted pricing function is a $(\mathbb{Q},\mathcal{F})$-martingale,
for $V\in C^{1,2}([0, T] \times \mathbb{R}^{p-1})$,
\begin{equation}\label{eq:martingale}
\mathbb{E}\left[dV(t, \boldsymbol{y}_t)\,\big|\,\mathcal F_t\right]
 = r(t, \boldsymbol{y}_t)V(t,\boldsymbol{y}_t)dt.
\end{equation}
Let $M^{(v)}(t, \boldsymbol{y}_t)$ denote the $v$-th column of
$M(t, \boldsymbol{y}_t)$. Applying It\^o's lemma to
$dV(t,\boldsymbol{y}_{t})$ and substituting into \eqref{eq:martingale} yields
the Feynman--Kac partial integro-differential equation:
\begin{align}\label{eq:FullFK}
 & \partial_{t}V+\boldsymbol{\mu}(t,\boldsymbol{y}_{t})^{\top}\nabla_{\boldsymbol{y}}V+\frac{1}{2}\text{tr}\left(\Sigma\Sigma^{\top}\nabla^2_{\boldsymbol{y}}V\right)\nonumber\\
 & -(r\left(t,\boldsymbol{y}_{t}\right)+\Lambda)V+\sum_{v=1}^{p-1}\lambda_{v}\,\mathbb E\left[V\left(t,\boldsymbol{y}_{t^-}+M^{(v)}\xi^{(v)}\right)|\mathcal{F}_t\right]=0 .
\end{align}
To lighten notation we write $\boldsymbol{x}=(t,\boldsymbol{y}_{t})$, a
$p$-vector.

The left-hand side of \eqref{eq:FullFK} decomposes into a local differential
operator and a non-local integral jump operator:
\begin{equation}\label{eq:LDeltaOps}
\left\{
\begin{aligned}
\mathcal{L}V(\boldsymbol x) &= \partial_t V -(r+\Lambda)V + \nabla_{\boldsymbol y} V^\top \boldsymbol{\mu} + \tfrac{1}{2} \text{tr}\left(\Sigma \Sigma^\top \nabla_{\boldsymbol y}^2 V\right), \\
\mathcal J V(\boldsymbol x) &= \sum_{v=1}^{p-1}\lambda_v \mathbb{E}\left[ V(t, \boldsymbol{y}_{t^-} + M^{(v)} \xi^{(v)})\,\middle|\,\mathcal{F}_t\right].
\end{aligned}
\right.
\end{equation}
This splitting is motivated by the fact that, for many jump-size
distributions, the expectation cannot be evaluated in closed form. Decoupling
the two operators allows the differential part to be computed analytically
inside the Gaussian process while isolating the non-local integral for
numerical treatment.

\section{Boundary value problem}
\label{sec:bvp}

Let
$\mathcal{Y}=[y^{(1)}_{\min}, y^{(1)}_{\max}]\times\cdots\times[y^{(p-1)}_{\min}, y^{(p-1)}_{\max}]$
denote the risk processes domain and $\partial \mathcal{Y}$  be its boundary.
Let $\mathring{\mathcal X}=(0,T)\times\mathcal Y$ the interior of the domain on
which the pricing function is defined.\footnote{For some derivatives the
natural domain is unbounded; the method requires a bounded one, so the spatial
bounds are placed far into the asymptotic tails, where the exit probability is
negligible, and an approximate asymptotic condition is imposed there.}
The set on which data is prescribed is the \emph{parabolic}
boundary
\begin{equation}\label{eq:parabbound}
  \partial\mathcal X
  := \underbrace{\bigl((0,T)\times\partial\mathcal Y\bigr)}_{=:\,\partial\mathcal X_{b}}
     \cup
     \underbrace{\bigl(\{T\}\times{\mathcal Y}\bigr)}_{=:\,\partial\mathcal X_{T}},
  \qquad
  \mathcal X:=\mathring{\mathcal X}\cup\partial\mathcal X .
\end{equation}
Note that $\partial\mathcal X$ is strictly smaller than the topological
boundary of $\mathcal X$: no condition is imposed on the initial face
$\{0\}\times\mathcal Y$, since a backward parabolic problem
carrying terminal data at $t=T$ would otherwise be over-determined.
On $\partial\mathcal X_{T}$ the price equals the payoff
$\mathcal P$; on $\partial\mathcal X_{b}$ it equals a known function $b$.

We infer that the pricing function of a European derivative with
payoff $\mathcal P$ solves
\begin{equation}\label{eq:R_prob}
\left\{
\begin{aligned}
\mathcal{L} V(\boldsymbol x) &= -\mathcal J V(\boldsymbol x) && \forall \boldsymbol x \in \mathring{\mathcal{X}}, \\
V(\boldsymbol x) &= h(\boldsymbol x) && \forall \boldsymbol x \in \partial \mathcal{X},
\end{aligned}
\right.
\end{equation}
where the boundary function $h$ is
\begin{eqnarray*}
h(\bar{\boldsymbol x}) & = & \begin{cases}
b(\bar{\boldsymbol x}) & \bar{\boldsymbol x}\in\partial\mathcal{X}_{b},\\
\mathcal{P}(\bar{\boldsymbol x}) & \bar{\boldsymbol x}\in\partial\mathcal{X}_{T}.
\end{cases}
\end{eqnarray*}

\section{The constrained Gaussian process regression (CGPR)}
\label{sec:cgpr}

Solving \eqref{eq:R_prob} can be viewed as a regression problem constrained by
a PIDE. We briefly review Gaussian process regression before introducing the
constrained extension, following \citet{gulian2019machine} and
\citet{chen2021solving}.

\subsection{Gaussian process}
A Gaussian process is a collection
$\{V(\boldsymbol{x}) : \boldsymbol{x}\in\mathcal{X}\}$ such that for any
finite $\boldsymbol{X} = \{\boldsymbol x_1, \dots, \boldsymbol x_N\} \subset \mathcal{X}$,
\begin{equation}\label{eq:GPf}
V(\boldsymbol{X})\sim \mathcal{N}\left(m(\boldsymbol{X}), k(\boldsymbol{X}, \boldsymbol{X})\right),
\end{equation}
with $k(\boldsymbol{X}, \boldsymbol{X})=\left(k(\boldsymbol{x}_{i},\boldsymbol{x}_{j})\right)_{i,j=1,...,N}$
and $k(\boldsymbol x,\tilde{\boldsymbol x})=\mathbb{C}\mathrm{ov}(V(\boldsymbol x), V(\tilde{\boldsymbol x}))$.
We use the Gaussian and Mat\'ern $5/2$ kernels:
\begin{equation}\label{eq:GandMKernels}
\left\{
\begin{aligned}
k^{G}(\boldsymbol x, \tilde{\boldsymbol x}) &=  a^2\exp \left( - \tfrac12\rho(\boldsymbol x, \tilde{\boldsymbol x})^2 \right), \\
k^{M}(\boldsymbol x, \tilde{\boldsymbol x}) &=  a^2\Big(1+\sqrt{5}\rho+\tfrac{5}{3}\rho^2\Big)\exp(-\sqrt{5}\rho),
\end{aligned}
\right.
\end{equation}
where $a^2>0$ is a signal variance and $\rho$ is the anisotropic distance
\begin{equation}\label{eq:rho_dist}
\rho(\boldsymbol x, \tilde{\boldsymbol x}) = \left(\sum_{i=1}^p \frac{(x_i-\tilde{x}_i)^2}{\eta_i^2}\right)^{1/2}.
\end{equation}
A necessary and sufficient condition for $k$ to be a valid kernel is that
$\left(k(\boldsymbol{x}_{i},\boldsymbol{x}_{j})\right)_{i,j}$ is positive
semi-definite for every finite sample. The mean function $m(\boldsymbol X)$ has little influence and can be set arbitrarily\footnote{In practice, the mean function corresponds to a prior belief on the solution. Setting it as a function close to the solution helps with convergence.}.

\subsection{Gaussian process regression}
Placing a Gaussian process prior on $V$ implies that the observations
$h(\boldsymbol{X})$ and the latent value $V(\boldsymbol x)$ are jointly Gaussian,
\begin{equation}\label{eq:jointDistributiongH}
\left(\begin{array}{c}
V(\boldsymbol x)\\
h(\boldsymbol{X})
\end{array}\right)\sim\mathcal{N}\left(\left(\begin{array}{c}
m(\boldsymbol x)\\
m(\boldsymbol{X})
\end{array}\right)\,,\,\left(\begin{array}{cc}
k(\boldsymbol{x},\boldsymbol{x}) & k(\boldsymbol{X},\boldsymbol{x})^\top\\
k(\boldsymbol{X},\boldsymbol{x}) & k(\boldsymbol{X},\boldsymbol{X})
\end{array}\right)\right),
\end{equation}
and the conditional expectation gives the estimator
\begin{eqnarray}\label{eq:mean_est}
\widehat V(\boldsymbol x) & = & m(\boldsymbol x)+k(\boldsymbol{X},\boldsymbol{x})^{\top}k(\boldsymbol{X},\boldsymbol{X})^{-1}\left(h(\boldsymbol{X})-m(\boldsymbol{X})\right),
\end{eqnarray}
with conditional variance
$k(\boldsymbol{x},\boldsymbol{x})-k(\boldsymbol{X},\boldsymbol{x})^{\top}k(\boldsymbol{X},\boldsymbol{X})^{-1}k(\boldsymbol{X},\boldsymbol{x})$.

\subsection{GPR constrained by the FK PIDE}

Estimating $V$ from a training set of prices produced by another method, as in
\citet{de2018machine}, is sub-optimal for two reasons. First, the
price function is completely determined by the PIDE together with the terminal
condition, so there is no need for exogenous prices. Second, such a regression
does not exploit \eqref{eq:martingale}, which is the functional constraint that
enforces absence of arbitrage. We therefore solve a regression constrained by
the equation itself.

We use two training sets: $n_B$ points on the boundary,
$\bar{\boldsymbol{X}}=\{\bar{\boldsymbol x}_{i}\}_{i=1,...,n_B}\subset\partial\mathcal{X}$,
with observations $h(\bar{\boldsymbol{X}})$; and $n_I$ interior collocation
points $\mathring{\boldsymbol{X}}=\{\mathring{\boldsymbol{x}}_{i}\}_{i=1,...,n_I}\subset\mathring{\mathcal{X}}$.
The core idea is that a linear operator maps a Gaussian process to a jointly
Gaussian process, so interior points can be incorporated directly alongside
boundary observations. For tractability we restrict the constraint to the local
operator $\mathcal{L}$, replacing $-\mathcal J(V)$ by an exogenous
source $z$ that does not depend on $V$ at the current iteration:
\begin{equation}\label{eq:R_prob_z}
\left\{
\begin{aligned}
\mathcal{L} V(\boldsymbol x) &= z(\boldsymbol x) && \forall \boldsymbol x \in \mathring{\mathcal{X}}, \\
V(\boldsymbol x) &= h(\boldsymbol x) && \forall \boldsymbol x \in \partial \mathcal{X}.
\end{aligned}
\right.
\end{equation}

The following result is standard for linear operators applied to Gaussian processes
\citep[Section~2]{sarkka2011linear}; see \citet{hainaut2024european} for the Feynman--Kac
operator.
\begin{proposition}\label{prop:GP_closure}
Let $V \sim \mathcal{GP}(m, k)$ on $\mathcal{X} \subset \mathbb{R}^{p}$, and
suppose the mixed partial derivatives of $k$ of order two in each argument
exist and are continuous on $\mathcal X\times\mathcal X$. Let
$\mathcal{L}_{\boldsymbol x}$ be the linear operator of
\eqref{eq:LDeltaOps}. Then $\mathcal{L}_{\boldsymbol x} V$ is a
well-defined mean-square Gaussian process, jointly Gaussian with $V$, with
\begin{align}
\mathbb{E}[\mathcal{L}_{\boldsymbol x} V(\boldsymbol x)] &= \mathcal{L}_{\boldsymbol x} m(\boldsymbol x), \\
\operatorname{Cov}(V(\boldsymbol x), \mathcal{L}_{\tilde{\boldsymbol x}} V(\tilde{\boldsymbol x})) &= \mathcal{L}_{\tilde{\boldsymbol x}} k(\boldsymbol x, \tilde{\boldsymbol x}), \\
\operatorname{Cov}(\mathcal{L}_{\boldsymbol x} V(\boldsymbol x), \mathcal{L}_{\tilde{\boldsymbol x}} V(\tilde{\boldsymbol x})) &= \mathcal{L}_{\boldsymbol x} \mathcal{L}_{\tilde{\boldsymbol x}} k(\boldsymbol x, \tilde{\boldsymbol x}).
\end{align}
\end{proposition}

\begin{remark}
A kernel from the Mat\'ern family with smoothness $\nu_{\mathrm M}$ is $2\lfloor\nu_{\mathrm M}\rfloor$
times differentiable at the origin, so $\nu_{\mathrm M}=5/2$ is the smallest member for
which $\mathcal L_{\boldsymbol x}\mathcal L_{\tilde{\boldsymbol x}}k$ exists.
The construction is therefore admissible but marginal for that kernel: the
radial functions appearing in Section~\ref{sec:maternkernel} are singular at
$\rho=0$ individually, and the diagonal of the Gram matrix must be evaluated
as a limit. Mat\'ern $3/2$ is not admissible.
\end{remark}

The joint distribution of a test evaluation and the training observations is
\begin{equation}\label{eq:CGPfullDist}
\begin{bmatrix}
V(\boldsymbol x) \\
V(\bar{\boldsymbol{X}}) \\
\mathcal{L} V(\mathring{\boldsymbol{X}})
\end{bmatrix}
\sim \mathcal{N} \left(
\begin{bmatrix}
m({\boldsymbol x})\\
m(\bar{\boldsymbol{X}})\\
\mathcal{L}m(\mathring{\boldsymbol{X}})
\end{bmatrix},
\begin{bmatrix}
k(\boldsymbol x, \boldsymbol x) & k(\bar{\boldsymbol{X}},\boldsymbol x)^\top & \bigl(\mathcal{L}_{\mathring{\boldsymbol x}} k(\mathring{\boldsymbol{X}},\boldsymbol x)\bigr)^\top \\
k(\bar{\boldsymbol{X}}, \boldsymbol x) & k(\bar{\boldsymbol{X}}, \bar{\boldsymbol{X}}) & \mathcal{L}_{\mathring{\boldsymbol x}} k(\bar{\boldsymbol{X}}, \mathring{\boldsymbol{X}})^\top \\
\mathcal{L}_{\mathring{\boldsymbol x}} k(\mathring{\boldsymbol{X}}, \boldsymbol x) & \mathcal{L}_{\mathring{\boldsymbol x}} k(\mathring{\boldsymbol{X}}, \bar{\boldsymbol{X}}) & \mathcal{L}_{\mathring{\boldsymbol x}} \mathcal{L}_{\tilde{\boldsymbol x}} k(\mathring{\boldsymbol{X}}, \mathring{\boldsymbol{X}})
\end{bmatrix}
\right)
\end{equation}
Conditioning on the known values
\begin{equation}\label{eq:trainingConditionConstrainedFKPDE}
\left\{ \begin{array}{llll}
V(\bar{\boldsymbol{x}}_{i}) & = & h_{i}, & i=1,\ldots,n_B,\\
\mathcal{L}_{\mathring{\boldsymbol x}}V(\mathring{\boldsymbol{x}}_{i}) & = & z_{i}, & i=1,\ldots,n_I,
\end{array}\right.
\end{equation}
gives the estimator.

\begin{proposition}\label{prop:DistPreds}
Define the Gram matrix and the cross-covariance vector
\begin{equation}\label{eq:Gram}
C(\mathring{\boldsymbol{X}}, \bar{\boldsymbol{X}})=
\begin{bmatrix}
 k(\bar{\boldsymbol{X}}, \bar{\boldsymbol{X}}) & \mathcal{L}_{\mathring{\boldsymbol x}} k(\mathring{\boldsymbol{X}}, \bar{\boldsymbol{X}})^\top \\
\mathcal{L}_{\mathring{\boldsymbol x}} k(\mathring{\boldsymbol{X}}, \bar{\boldsymbol{X}}) & \mathcal{L}_{\mathring{\boldsymbol x}} \mathcal{L}_{\tilde{\boldsymbol x}} k(\mathring{\boldsymbol{X}}, \mathring{\boldsymbol{X}})
\end{bmatrix},
\qquad
\boldsymbol{\beta}(\boldsymbol x)=
\begin{bmatrix}
k(\bar{\boldsymbol{X}}, \boldsymbol x)\\
\mathcal{L}_{\mathring{\boldsymbol x}} k(\mathring{\boldsymbol{X}}, \boldsymbol x)
\end{bmatrix}.
\end{equation}
Then $V(\boldsymbol x)$ conditioned on the training data is Gaussian with
\begin{equation}\label{eq:ExPreds}
 \widehat V(\boldsymbol x) = m(\boldsymbol x) + \boldsymbol{\beta}(\boldsymbol x)^\top
C(\mathring{\boldsymbol{X}}, \bar{\boldsymbol{X}})^{-1}
\begin{bmatrix}
h(\bar{\boldsymbol{X}}) - m(\bar{\boldsymbol{X}})\\
z(\mathring{\boldsymbol{X}}) - \mathcal{L} m(\mathring{\boldsymbol{X}})
\end{bmatrix}
\end{equation}
and
\begin{align}\label{eq:CovPreds}
\mathbb V[V(\boldsymbol x)] =& k(\boldsymbol x, \boldsymbol x)-
\boldsymbol{\beta}(\boldsymbol x)^\top
C(\mathring{\boldsymbol{X}}, \bar{\boldsymbol{X}})^{-1}
\boldsymbol{\beta}(\boldsymbol x).
\end{align}
\end{proposition}
\begin{proof}
By Proposition~\ref{prop:GP_closure}, the vector
$\bigl(V(\boldsymbol x),V(\bar{\boldsymbol X}),\mathcal LV(\mathring{\boldsymbol X})\bigr)$
is jointly Gaussian with mean and covariance given by \eqref{eq:CGPfullDist}.
The result follows from the conditional distribution of a partitioned Gaussian
vector, as for noise-free observations \citep[Chap.~2]{rasmussen2006gaussian}; see also
\citet{sarkka2011linear} for observations of linear operators of the process.
\end{proof}
The Gram matrix is symmetric. The block
$k(\bar{\boldsymbol X},\bar{\boldsymbol X})$ is symmetric because $k$ is;
the operator acting on the first argument coincides with the operator acting on
the second argument of the reversed kernel, so the off-diagonal blocks are
mutual transposes; and since $\boldsymbol x$ and $\tilde{\boldsymbol x}$ are
independent arguments the two operators commute, giving
$\mathcal{L}_{\boldsymbol x} \mathcal{L}_{\tilde{\boldsymbol x}} k(\boldsymbol x_i, \boldsymbol x_j)
= \mathcal{L}_{\tilde{\boldsymbol x}} \mathcal{L}_{\boldsymbol x} k(\boldsymbol x_j, \boldsymbol x_i)$.

\section{Fixed-point method and convergence}
\label{sec:fixedpoint}

Section~\ref{sec:cgpr} provides a regression method able to solve local boundary value
problems of the form $\mathcal L V=z$. The pricing problem \eqref{eq:R_prob},
however, has a right-hand side $-\mathcal J(V)$ that depends on the solution.
Following the splitting idea of \citet{hainaut2026american}, we treat the jump
term as a lagged source: starting from an initial source $z^{(0)}$, we iterate
\begin{equation}\label{eq:iteration}
  \left\{
  \begin{aligned}
    \mathcal{L}V^{(n)} &= z^{(n-1)} && \text{on }\mathring{\mathcal X}\\
    V^{(n)} &= h && \text{on }\partial\mathcal X\\
    z^{(n)} &= -\mathcal J\bigl(V^{(n)}\bigr) &&
  \end{aligned}
  \right.
  \qquad n\ge1
\end{equation}
The first two lines form a local problem of the type \eqref{eq:R_prob_z}, solved
in closed form by the constrained regression of Section~\ref{sec:cgpr}; the third
evaluates the jump expectation on the new price surface. This section formalises
the loop and establishes its convergence.

\subsection{Function space and extension of the jump operator}

Let $\mathcal B:=\bigl(C_b(\mathcal X),\|\cdot\|_{\infty}\bigr)$ be the Banach space of
bounded continuous functions on $\mathcal X$ under the supremum norm, write
$M^{(v)}(t,\boldsymbol y):=M(t,\boldsymbol y)\boldsymbol e_{v}$ for the $v$-th column
of $M$, and $\pi_{v}$ for the law of $\xi^{(v)}$, a probability measure on $\mathbb R$.
The operator $\mathcal J$ of \eqref{eq:LDeltaOps} evaluates $V$ at the post-jump states
$\boldsymbol y+M^{(v)}(t,\boldsymbol y)\xi$, which may lie outside $\mathcal Y$, so it is
not a priori defined on $\mathcal B$. Outside the domain, the value of the claim is
fixed by the contract or by the truncation convention, not by the solution: for a
knock-out barrier it is the rebate, and for a vanilla claim on a truncated domain it
is the asymptotic value already used on the boundary, which for a call grows without
bound. We therefore suppose that the boundary function $h$ is defined on $[0,T]\times(\mathbb R^{p-1}\setminus\mathcal Y)$, and every candidate
solution $V\in\mathcal B$ is understood to satisfy
\begin{equation}\label{eq:extension}
  V(t,\boldsymbol y')=h(t,\boldsymbol y'),
  \qquad \boldsymbol y'\notin\mathcal Y .
\end{equation}
Using the integral formulation of the expectation in \eqref{eq:LDeltaOps}:
\begin{equation}\label{eq:Delta-ext}
  \mathcal J(V)(t,\boldsymbol y)
  :=\sum_{v=1}^{p-1}\lambda_{v}\int_{\mathbb R}
    V\bigl(t,\boldsymbol y+M^{(v)}(t,\boldsymbol y)\,\xi\bigr)\,\pi_{v}(d\xi),
\end{equation}
which is the meaning of $\mathcal J$ throughout, including in \eqref{eq:R_prob} and
\eqref{eq:iteration}.

\subsection{Preliminaries}
We now state the Weissinger theorem \citep{weissinger1952theorie}, which is central to the convergence demonstration. It will be used to show that our splitting method, combined with the loop we built, converges to the derivative price with jumps.

\begin{theorem}[Weissinger's fixed-point theorem]\label{thm:weissinger}
Let $(X,\|\cdot\|)$ be a Banach space, $\mathcal S\subseteq X$ non-empty and
closed, and $\mathcal T:\mathcal S\to\mathcal S$. Suppose there exist $(\omega_{n})_{n\geq1}\ge0$
with $\sum_{n\ge1}\omega_{n}<\infty$ and
\begin{equation}\label{eq:weissinger-hyp}
  \bigl\|\mathcal{T}^{n}(x)-\mathcal{T}^{n}(y)\bigr\|\le \omega_{n}\|x-y\|
  \qquad\forall x,y\in\mathcal S,\ \forall n\ge1 .
\end{equation}
Then $\mathcal T$ has a unique fixed point $x^{*}\in\mathcal S$; for every
$x_{0}\in\mathcal S$ the sequence $x_{n}:=\mathcal T^{n}(x_{0})$ converges to
$x^{*}$; and
\begin{equation}\label{eq:weissinger-err}
  \|x^{*}-x_{n}\| \le \Bigl(\textstyle\sum_{k\ge n}\omega_{k}\Bigr)\|x_{1}-x_{0}\| .
\end{equation}
\end{theorem}

We also state the Feynman--Kac representation for boundary value
problems \citep{friedman1975stochastic}, applied to a jump-free diffusion
\begin{equation}\label{eq:aux-sde}
  d\widetilde{\boldsymbol Y}_{s}
  = \boldsymbol\mu(s,\widetilde{\boldsymbol Y}_{s})\,ds
  + \Sigma(s,\widetilde{\boldsymbol Y}_{s})\,d\boldsymbol W_{s},
\end{equation}
obtained from \eqref{eq:JDmodel} by suppressing the jump component while
retaining the risk-neutral drift, and discounted at the rate $r+\Lambda$ carried by
$\mathcal L$.

We also recall the Feynman--Kac Theorem for the BVP problem without jumps. Its proof is given in \citet{friedman1975stochastic}.
\begin{theorem}[Feynman--Kac formula for terminal/boundary value problems]\label{thm:feynman_kac_bvp}
Let $h$ be a continuous function on $\partial \mathcal X$, and $\phi$ be a Hölder continuous function on $\mathcal X$. Let $\tau_{\mathcal X}$ be the exit time of $\widetilde{\boldsymbol Y}_{t}$ from $\mathcal X$.
There exists a unique function $u\in C^{1,2}(\mathcal X)$ solving:
\begin{equation}\label{eq:fk_pde}
\left\{
\begin{aligned}
&\partial_t u+\boldsymbol\mu^{\top}\nabla_{\boldsymbol y}u
 +\tfrac12\mathrm{tr}\bigl(\Sigma\Sigma^{\top}\nabla^2_{\boldsymbol y}u\bigr)
 -(r+\Lambda)\,u+\phi=0 && \text{on }\mathring{\mathcal X}\\
&u=h && \text{on }\partial\mathcal X
\end{aligned}
\right.
\end{equation}
that admits the stochastic representation:
\begin{equation}\label{eq:fk_solution}
\begin{aligned}
    u(t, \boldsymbol{y}) =\;& \mathbb{E}_{t, \boldsymbol{y}} \left[ e^{-\int_t^{\tau_{\mathcal X}} (r+\Lambda)(s, \widetilde{\boldsymbol{Y}}_s)ds} h(\tau_{\mathcal X}, \widetilde{\boldsymbol{Y}}_{\tau_{\mathcal X}})\right] \\
&+ \mathbb{E}_{t, \boldsymbol{y}} \left[\int_t^{\tau_{\mathcal X}} e^{-\int_t^s (r+\Lambda)(v, \widetilde{\boldsymbol{Y}}_v)dv} \phi(s, \widetilde{\boldsymbol{Y}}_s)\, ds \right]
\end{aligned}
\end{equation}
\end{theorem}

\subsection{Convergence of the iteration}

We assume throughout:
\begin{itemize}
\item[\textbf{H1}] $\boldsymbol\mu$, $\Sigma$, $M$ and $r$ are bounded and
  continuous on $\mathcal X$,
  and
  \begin{equation}\label{eq:b-def}
    \beta_0:=r_{\min}+\Lambda>0,\qquad r_{\min}:=\inf_{\mathcal X}r(t, \boldsymbol{y}_t) .
  \end{equation}
  Note that since $\Lambda>0$, condition \eqref{eq:b-def} holds automatically when $r\ge0$.
\item[\textbf{H2}] For every $z\in\mathcal B$, the local problem
  \eqref{eq:R_prob_z} admits a solution in $\mathcal B$ that is classical on
  $\mathring{\mathcal X}$.

\end{itemize}

By the last assertion of Theorem~\ref{thm:feynman_kac_bvp},
the solution in \textbf{H2} is unique. Each step of
\eqref{eq:iteration} is therefore well defined: $V^{(n)}$ is the unique solution
of \eqref{eq:R_prob_z} with source $z^{(n-1)}$, and $z^{(n)}=-\mathcal J(V^{(n)})$.
The map $z^{(n-1)}\mapsto z^{(n)}$ plays the role of $\mathcal T$ in
Theorem~\ref{thm:weissinger}, and $z^{(n)}$ is its $n$-th iterate from $z^{(0)}$.
A fixed point is a source $z^{*}$ that reproduces its own jump term,
$z^{*}=-\mathcal J(V^{*})$, where $V^{*}$ solves \eqref{eq:R_prob_z} with source
$z^{*}$; this $V^{*}$ then solves \eqref{eq:R_prob}, as the end of the proof of
Proposition~\ref{prop:weiss-main} verifies. When two iterations are compared, we
write $z_{i}^{(n)}$ and $V_{i}^{(n)}$, $i=1,2$, for the iterates of
\eqref{eq:iteration} started from $z_{i}^{(0)}$.

\begin{lemma}[The iteration stays in $\mathcal B$]\label{lem:selfmap}
Under \textbf{H1}--\textbf{H2}, for every $V\in\mathcal B$,
\begin{equation}\label{eq:Delta-bound}
  \bigl\|\mathcal J(V)\bigr\|_{\infty}\;\le\;\Lambda\|V\|_{\infty}
\end{equation}
and $\mathcal J(\mathcal B)\subseteq\mathcal B$. In particular, $z^{(n)}\in\mathcal B$
for all $n\ge0$ whenever $z^{(0)}\in\mathcal B$.
\end{lemma}

\begin{proof}
Fix $V\in\mathcal B$. For $(t,\boldsymbol y)\in\mathcal X$ and $\xi\in\mathbb R$,
the post-jump value is bounded by $\|V\|_{\infty}$. Hence:
\begin{equation}\label{eq:ext-dom}
  \bigl|V\bigl(t,\boldsymbol y+M^{(v)}(t,\boldsymbol y)\xi\bigr)\bigr|
  \;\le\;\|V\|_{\infty}
\end{equation}

For $(t,\boldsymbol y)\in\mathcal X$, by the triangle
inequality, \eqref{eq:ext-dom} and $\pi_{v}(\mathbb R)=1$,
\begin{align}
  \bigl|\mathcal J(V)(t,\boldsymbol y)\bigr|
  &\le\sum_{v=1}^{p-1}\lambda_{v}\int_{\mathbb R}
     \bigl|V\bigl(t,\boldsymbol y+M^{(v)}(t,\boldsymbol y)\xi\bigr)\bigr|\,\pi_{v}(d\xi)
  \notag\\
  &\le\sum_{v=1}^{p-1}\lambda_{v}\int_{\mathbb R}
     \|V\|_{\infty}\,\pi_{v}(d\xi)\notag\\
  &=\;\Lambda\|V\|_{\infty}
  \label{eq:selfmap-bdd}
\end{align}

Finally, if $z^{(n-1)}\in\mathcal B$, \textbf{H2} gives
$V^{(n)}\in\mathcal B$, hence $z^{(n)}=-\mathcal J(V^{(n)})\in\mathcal B$ because
$\mathcal B$ is a linear space, and the claim follows by induction.
\end{proof}

\begin{lemma}[Lipschitz continuity of $\mathcal J$]\label{lem:D-lip}
Under \textbf{H1}--\textbf{H2}, for all $V_{1},V_{2}\in\mathcal B$ and all
$t\in[0,T]$,
\begin{equation}\label{eq:D-lip}
  \sup_{\boldsymbol y\in\mathcal Y}\bigl|\mathcal J(V_{1})-\mathcal J(V_{2})\bigr|(t,\boldsymbol y)
  \;\le\;\Lambda\,\sup_{\boldsymbol y\in\mathcal Y}\bigl|V_{1}-V_{2}\bigr|(t,\boldsymbol y).
\end{equation}
\end{lemma}

\begin{proof}
Fix $(t,\boldsymbol y)\in\mathcal X$ and two pricing functions $V_1$ and $V_2$, solution of \eqref{eq:R_prob_z}. As these functions share the same values $h(t, \boldsymbol y_t)$ outside $\mathcal{X}$,
\begin{equation}\label{eq:ext-slice}
  \sup_{\boldsymbol y'\in\mathbb R^{p-1}}\bigl|V_{1}-V_{2}\bigr|(t,\boldsymbol y')
  =\sup_{\boldsymbol y\in\mathcal Y}\bigl|V_{1}-V_{2}\bigr|(t,\boldsymbol y),
\end{equation}

The integrals defining $\mathcal J(V_{1})$ and
$\mathcal J(V_{2})$ are finite by Lemma~\ref{lem:selfmap}, so they may be subtracted
under the integral sign. \eqref{eq:ext-slice} with $\pi_{v}(\mathbb R)=1$ gives:
\begin{align*}
  \bigl|\mathcal J(V_{1})-\mathcal J(V_{2})\bigr|(t,\boldsymbol y)
  &=\Bigl|\sum_{v=1}^{p-1}\lambda_{v}\int_{\mathbb R}
     (V_{1}-V_{2})(t,\boldsymbol y+M^{(v)}(t,\boldsymbol y)\xi)\,\pi_{v}(d\xi)\Bigr|\\
  &\le\sum_{v=1}^{p-1}\lambda_{v}\int_{\mathbb R}
     \bigl|V_{1}-V_{2}\bigr|(t,\boldsymbol y+M^{(v)}(t,\boldsymbol y)\xi)\,\pi_{v}(d\xi)\\
  &\le\sum_{v=1}^{p-1}\lambda_{v}\,
     \sup_{\boldsymbol y'\in\mathbb R^{p-1}}\bigl|V_{1}-V_{2}\bigr|(t,\boldsymbol y')\,
     \pi_{v}(\mathbb R)\\
  &=\Lambda\sup_{\boldsymbol y'\in\mathcal Y}\bigl|V_{1}-V_{2}\bigr|(t,\boldsymbol y').
\end{align*}
The right-hand side does not depend on $\boldsymbol y$, so taking the supremum
over $\boldsymbol y\in\mathcal Y$ on the left gives \eqref{eq:D-lip}.
\end{proof}

\begin{lemma}[Time-dependent bound on the local solve]\label{lem:G-graded}
Under \textbf{H1}--\textbf{H2}, let $z_{1},z_{2}\in\mathcal B$, let
$V_{1},V_{2}\in\mathcal B$ be the solutions of \eqref{eq:R_prob_z} with sources
$z_{1}$ and $z_{2}$, and define the time-dependent supremum:
\begin{equation}\label{eq:profile}
  E(s):=\sup_{\boldsymbol y\in\mathcal Y}\bigl|z_{1}-z_{2}\bigr|(s,\boldsymbol y),
  \qquad s\in[0,T].
\end{equation}
Then, for all $(t,\boldsymbol y)\in\mathcal X$,
\begin{equation}\label{eq:G-graded}
  \bigl|V_{1}-V_{2}\bigr|(t,\boldsymbol y)
  \;\le\;\int_{t}^{T} e^{-\beta_0(s-t)}\,E(s)\,ds .
\end{equation}
\end{lemma}

\begin{proof}

Set $u:=V_{1}-V_{2}\in\mathcal B$. By \textbf{H2} and
linearity of $\mathcal L$,
\begin{equation*}
  \mathcal{L}u=z_{1}-z_{2}\ \text{ on }\mathring{\mathcal X},
  \qquad u=h-h=0\ \text{ on }\partial\mathcal X .
\end{equation*}
Since $\mathcal L$ discounts at $r+\Lambda$ by \eqref{eq:LDeltaOps}, this is
\eqref{eq:fk_pde} with
\begin{equation*}
  \phi=z_{2}-z_{1},\qquad h\equiv0 ,
\end{equation*}
and Theorem~\ref{thm:feynman_kac_bvp} can be applied and gives:
\begin{equation}\label{eq:u-repr}
  u(t,\boldsymbol y)=\mathbb{E}_{t,\boldsymbol y}\left[
    \int_{t}^{\tau_{\mathcal X}}
      e^{-\int_{t}^{s}(r+\Lambda)(v,\widetilde{\boldsymbol Y}_{v})\,dv}
      \bigl(z_{2}-z_{1}\bigr)(s,\widetilde{\boldsymbol Y}_{s})\,ds\right].
\end{equation}
Hence
\begin{align}
  \bigl|u(t,\boldsymbol y)\bigr|
  &\le\mathbb{E}_{t,\boldsymbol y}\left[\int_{t}^{\tau_{\mathcal X}}
     e^{-\int_{t}^{s}(r+\Lambda)(v,\widetilde{\boldsymbol Y}_{v})\,dv}
     \bigl|z_{1}-z_{2}\bigr|(s,\widetilde{\boldsymbol Y}_{s})\,ds\right]
  \\
  &\le\mathbb{E}_{t,\boldsymbol y}\left[\int_{t}^{\tau_{\mathcal X}}
     e^{-\beta_0(s-t)}\bigl|z_{1}-z_{2}\bigr|(s,\widetilde{\boldsymbol Y}_{s})\,ds\right]\\
  &\le\mathbb{E}_{t,\boldsymbol y}\left[\int_{t}^{\tau_{\mathcal X}}
     e^{-\beta_0(s-t)}E(s)\,ds\right]\\
  &\le\mathbb{E}_{t,\boldsymbol y}\left[\int_{t}^{T}
     e^{-\beta_0(s-t)}E(s)\,ds\right]\\
  &=\int_{t}^{T}e^{-\beta_0(s-t)}E(s)\,ds
  \label{eq:G-chain}
\end{align}
\end{proof}

\begin{proposition}[Iteration convergence]\label{prop:weiss-main}
Under \textbf{H1}--\textbf{H2}, for all $z_{1}^{(0)},z_{2}^{(0)}\in\mathcal B$ and $n\ge1$,
\begin{equation}\label{eq:superlinear}
  \bigl\|z_{1}^{(n)}-z_{2}^{(n)}\bigr\|_{\infty}
  \;\le\;\frac{(\Lambda T)^{n}}{n!}\,\bigl\|z_{1}^{(0)}-z_{2}^{(0)}\bigr\|_{\infty}.
\end{equation}
Consequently the iteration \eqref{eq:iteration} admits a unique fixed point
$z^{*}\in\mathcal B$ and converges to it from any $z^{(0)}\in\mathcal B$, and the
solution $V^{*}$ of \eqref{eq:R_prob_z} with source $z^{*}$ is the unique solution
of \eqref{eq:R_prob} in $\mathcal B$ that is classical on $\mathring{\mathcal X}$.
Moreover
\begin{alignat}{2}\label{eq:weiss-apriori}
  &\bigl\|z^{*}-z^{(n)}\bigr\|_{\infty} &&\le\Bigl(\sum_{k\ge n}\tfrac{(\Lambda T)^{k}}{k!}\Bigr)\bigl\|z^{(1)}-z^{(0)}\bigr\|_{\infty} \notag\\
  &\bigl\|V^{*}-V^{(n)}\bigr\|_{\infty}
  &&\le\Bigl(\frac{1-e^{-\beta_0T}}{\beta_0}\Bigr)\bigl\|z^{*}-z^{(n-1)}\bigr\|_{\infty}
\end{alignat}

\end{proposition}

\begin{proof}
Let $z_{1}^{(0)},z_{2}^{(0)}\in\mathcal B$, let
$\delta_{0}:=\|z_{1}^{(0)}-z_{2}^{(0)}\|_{\infty}$, and set, for $n\ge0$,
\begin{equation}\label{eq:En-def}
  E_{n}(t):=\sup_{\boldsymbol y\in\mathcal Y}
            \bigl|z_{1}^{(n)}-z_{2}^{(n)}\bigr|(t,\boldsymbol y),
  \qquad t\in[0,T],
\end{equation}
so that $E_{0}(t)\le\delta_{0}$. By Lemma~\ref{lem:selfmap},
$z_{i}^{(n)},V_{i}^{(n)}\in\mathcal B$ for all $n$. By \eqref{eq:iteration},
$z_{i}^{(n)}=-\mathcal J\bigl(V_{i}^{(n)}\bigr)$, hence, pointwise on $\mathcal X$,
\begin{equation*}
  \bigl|z_{1}^{(n)}-z_{2}^{(n)}\bigr|
  =\Bigl|\mathcal J\bigl(V_{1}^{(n)}\bigr)-\mathcal J\bigl(V_{2}^{(n)}\bigr)\Bigr| ,
\end{equation*}
and Lemma~\ref{lem:D-lip} gives
\begin{equation}\label{eq:step1}
  E_{n}(t)
  \;\le\;\Lambda\,\sup_{\boldsymbol y\in\mathcal Y}
    \bigl|V_{1}^{(n)}-V_{2}^{(n)}\bigr|(t,\boldsymbol y),
  \qquad n\ge1,\ t\in[0,T].
\end{equation}

Apply Lemma~\ref{lem:G-graded} to $V_{1}^{(n)},V_{2}^{(n)}$, the solutions with
sources $z_{1}^{(n-1)},z_{2}^{(n-1)}\in\mathcal B$. By \eqref{eq:En-def} their
profile \eqref{eq:profile} is $E_{n-1}$, so for all $(t,\boldsymbol y)\in\mathcal X$
\begin{equation}\label{eq:step2a}
  \bigl|V_{1}^{(n)}-V_{2}^{(n)}\bigr|(t,\boldsymbol y)
  \;\le\;\int_{t}^{T}e^{-\beta_0(s-t)}E_{n-1}(s)\,ds .
\end{equation}
The right-hand side does not depend on $\boldsymbol y$, so it also bounds the
supremum over $\boldsymbol y\in\mathcal Y$ of the left-hand side. Inserting this
into \eqref{eq:step1},
\begin{equation}\label{eq:volterra}
  E_{n}(t)\;\le\;\Lambda\int_{t}^{T}e^{-\beta_0(s-t)}\,E_{n-1}(s)\,ds,
  \qquad n\ge1,\ t\in[0,T].
\end{equation}

By \eqref{eq:b-def}, $\beta_0>0$, so
$e^{-\beta_0(s-t)}\le1$ for $s\ge t$ and \eqref{eq:volterra} implies
\begin{equation}\label{eq:gronwall-step}
  E_{n}(t)\;\le\;\Lambda\int_{t}^{T}E_{n-1}(s)\,ds,
  \qquad n\ge1,\ t\in[0,T].
\end{equation}

For the following, we proceed by induction. We claim
\begin{equation}\label{eq:induction}
  E_{n}(t)\;\le\;\frac{\Lambda^{n}(T-t)^{n}}{n!}\,\delta_{0},
  \qquad n\ge0,\ t\in[0,T].
\end{equation}
For $n=0$ this is $E_{0}(t)\le\delta_{0}$. If \eqref{eq:induction} holds at
order $n-1$, then by \eqref{eq:gronwall-step}
\begin{equation*}
  E_{n}(t)
  \le\Lambda\int_{t}^{T}\frac{\Lambda^{n-1}(T-s)^{n-1}}{(n-1)!}\,\delta_{0}\,ds
  =\frac{\Lambda^{n}\delta_{0}}{(n-1)!}\cdot\frac{(T-t)^{n}}{n}
  =\frac{\Lambda^{n}(T-t)^{n}}{n!}\,\delta_{0},
\end{equation*}
which is \eqref{eq:induction} at order $n$.
Finally:
\begin{equation*}
  \bigl\|z_{1}^{(n)}-z_{2}^{(n)}\bigr\|_{\infty}
  =\sup_{t\in[0,T]}E_{n}(t)
  \le\sup_{t\in[0,T]}\frac{\Lambda^{n}(T-t)^{n}}{n!}\,\delta_{0}
  =\frac{(\Lambda T)^{n}}{n!}\,\bigl\|z_{1}^{(0)}-z_{2}^{(0)}\bigr\|_{\infty},
\end{equation*}
which is \eqref{eq:superlinear}.

We can finally apply Weissinger by setting $\omega_{n}:=(\Lambda T)^{n}/n!$. These
constants do not depend on $z_{1}^{(0)},z_{2}^{(0)}$, and
\begin{equation}\label{eq:summable}
  \sum_{n\ge1}\omega_{n}=e^{\Lambda T}-1<\infty .
\end{equation}
By Lemma~\ref{lem:selfmap}, \eqref{eq:superlinear} and \eqref{eq:summable}, the
hypotheses of Theorem~\ref{thm:weissinger} hold with
$\mathcal S=X=\mathcal B$ and $\mathcal T$ the map $z^{(n-1)}\mapsto z^{(n)}$.
Hence the iteration has a unique fixed point $z^{*}\in\mathcal B$, and for every
$z^{(0)}\in\mathcal B$ the sequence $z^{(n)}$ of \eqref{eq:iteration} converges to
$z^{*}$ with, by \eqref{eq:weissinger-err},
\begin{equation*}
  \bigl\|z^{*}-z^{(n)}\bigr\|_{\infty}
  \le\Bigl(\sum_{k\ge n}\omega_{k}\Bigr)\bigl\|z^{(1)}-z^{(0)}\bigr\|_{\infty}.
\end{equation*}
For the price, $V^{*}$ and $V^{(n)}$ solve \eqref{eq:R_prob_z} with sources
$z^{*}$ and $z^{(n-1)}$, and the profile \eqref{eq:profile} of the pair $(z^{*},z^{(n-1)})$ is bounded by
$\|z^{*}-z^{(n-1)}\|_{\infty}$. Lemma~\ref{lem:G-graded} gives, for
$(t,\boldsymbol y)\in\mathcal X$,
\begin{align}\label{eq:err-V}
  \bigl|V^{*}-V^{(n)}\bigr|(t,\boldsymbol y)
  &\le\|z^{*}-z^{(n-1)}\|_{\infty}\int_{t}^{T}e^{-\beta_0(s-t)}\,ds\notag\\
  &=\|z^{*}-z^{(n-1)}\|_{\infty}\,\frac{1-e^{-\beta_0(T-t)}}{\beta_0}\notag\\
  &\le\|z^{*}-z^{(n-1)}\|_{\infty}\,\frac{1-e^{-\beta_0T}}{\beta_0}.
\end{align}
Taking the supremum over
$(t,\boldsymbol y)\in\mathcal X$ gives the second bound in
\eqref{eq:weiss-apriori}.

The last step required is to show that $V^*$ obtained after convergence is the price. By definition of $V^{*}$ and since $z^{*}$ is a fixed point,
\begin{equation*}
  \mathcal LV^{*}=z^{*}\ \text{ on }\mathring{\mathcal X},
  \qquad V^{*}=h\ \text{ on }\partial\mathcal X,
  \qquad z^{*}=-\mathcal J\bigl(V^{*}\bigr).
\end{equation*}
Substituting the third identity into the first,
\begin{equation}\label{eq:Vstar}
  \mathcal LV^{*}=-\mathcal J\bigl(V^{*}\bigr)\ \text{ on }\mathring{\mathcal X},
  \qquad V^{*}=h\ \text{ on }\partial\mathcal X,
\end{equation}
which is \eqref{eq:R_prob}; and $V^{*}\in\mathcal B$ is classical on
$\mathring{\mathcal X}$ by \textbf{H2}.

Conversely, let $V\in\mathcal B$, classical on $\mathring{\mathcal X}$, solve
\eqref{eq:R_prob}, and set $z:=-\mathcal J(V)$, which lies in $\mathcal B$ by
Lemma~\ref{lem:selfmap}. Then
\begin{equation*}
  \mathcal LV=z\ \text{ on }\mathring{\mathcal X},\qquad V=h\ \text{ on }\partial\mathcal X,
\end{equation*}
so $V$ is, by uniqueness, the solution of \eqref{eq:R_prob_z} with source $z$,
and one step of \eqref{eq:iteration} started from $z$ returns $-\mathcal J(V)=z$:
$z$ is a fixed point. Uniqueness of the fixed point gives $z=z^{*}$, and
uniqueness of the solution of \eqref{eq:R_prob_z} with source $z^{*}$ gives
$V=V^{*}$.
\end{proof}

\subsection{Stopping Criterion}
\label{sec:stopping}

Proposition~\ref{prop:weiss-main} gives more than convergence. It bounds the
error of the $n$-th iterate by a quantity that depends only on $n$ and on the
first increment of the iteration, and this bound can be inverted to obtain a
number of iterations that guarantees any prescribed precision. Throughout,
write
\begin{equation*}
  c_T:=\frac{1-e^{-\beta_0T}}{\beta_0},\qquad
  (\omega_{k})_{k\geq0}:=\frac{(\Lambda T)^{k}}{k!},\qquad
  U_{j}:=\sum_{k\ge j}\omega_{k},
\end{equation*}
so that $U_{j}$ is the tail of the exponential series and
$U_{0}=e^{\Lambda T}$.

\begin{proposition}[Worst case convergence for a given precision]\label{prop:stopping}
Under the assumptions of Proposition~\ref{prop:weiss-main}, for every $n\ge1$,
\begin{equation}\label{eq:apriori-V}
  \bigl\|V^{*}-V^{(n)}\bigr\|_{\infty}\;\le\;c_T\,U_{n-1}\,\|z^{(1)}-z^{(0)}\|_{\infty}.
\end{equation}
If the iteration is started from $z^{(0)}\equiv0$, then
\begin{equation}\label{eq:delta1}
  \|z^{(1)}-z^{(0)}\|_{\infty}\;\le\;\Lambda\,\|h\|_{\infty},
\end{equation}
so that the right-hand side of \eqref{eq:apriori-V} is known before the first
iteration. Consequently, for every $\varepsilon>0$, the integer
\begin{equation}\label{eq:Neps}
  N(\varepsilon):=\min\bigl\{n\ge1:\ c_T\,U_{n-1}\,\|z^{(1)}-z^{(0)}\|_{\infty}\le\varepsilon\bigr\}
\end{equation}
satisfies $\|V^{*}-V^{(n)}\|_{\infty}\le\varepsilon$ for all
$n\ge N(\varepsilon)$.
\end{proposition}

\begin{proof}
The iterates $z^{(k+1)}$ and $z^{(k)}$ are the $k$-th iterates of
\eqref{eq:iteration} started from $z^{(1)}$ and $z^{(0)}$, so
\eqref{eq:superlinear} gives
$\|z^{(k+1)}-z^{(k)}\|_{\infty}\le \omega_{k}\|z^{(1)}-z^{(0)}\|_{\infty}$, which also
holds for $k=0$ since $\omega_{0}=1$. Writing $z^{*}-z^{(j)}$ as a telescoping series,
\begin{align*}
  \bigl\|z^{*}-z^{(j)}\bigr\|_{\infty}
  &\le \sum_{k\ge j}\bigl\|z^{(k+1)}-z^{(k)}\bigr\|_{\infty} \\
  &\le \sum_{k\ge j}\omega_{k}\,\|z^{(1)}-z^{(0)}\|_{\infty} \\
  &= U_{j}\,\|z^{(1)}-z^{(0)}\|_{\infty}.
\end{align*}
Combining this with the second bound in \eqref{eq:err-V},
$\|V^{*}-V^{(n)}\|_{\infty}\le c_T\|z^{*}-z^{(n-1)}\|_{\infty}$, gives
\eqref{eq:apriori-V}.

If $z^{(0)}\equiv0$, then $z^{(1)}=-\mathcal J(V^{(1)})$, where $V^{(1)}$ solves
\eqref{eq:R_prob_z} with zero source. By Theorem~\ref{thm:feynman_kac_bvp} with
$\phi=0$ and discount rate $r+\Lambda\ge \beta_0>0$, $V^{(1)}$ is a discounted
expectation of $h$, so $\|V^{(1)}\|_{\infty}\le\|h\|_{\infty}$, and
\eqref{eq:Delta-bound} gives
$\|z^{(1)}-z^{(0)}\|_{\infty}=\|\mathcal J(V^{(1)})\|_{\infty}
\le\Lambda\|h\|_{\infty}$, which is \eqref{eq:delta1}.

Since every $\omega_{k}$ is non-negative, $U_{j}$ is non-increasing in $j$, so once
$c_T U_{n-1}\|z^{(1)}-z^{(0)}\|_{\infty}\le\varepsilon$ holds it holds for all
larger $n$.
\end{proof}

Proposition~\ref{prop:stopping} provides an a priori stopping criterion that guarantees convergence within any precision $\varepsilon$ without evaluating the uniform norm $\Vert{}\cdot\Vert{}_\infty$. This is essential in practice, as the true supremum norm across the continuous domain cannot be evaluated directly from a finite set of collocation points.

\subsection{Numerical implementation}
\label{sec:implementation}

The estimator of Proposition~\ref{prop:DistPreds} requires solving linear
systems with the dense Gram matrix $C(\mathring{\boldsymbol X},\bar{\boldsymbol X})$.
Kernel matrices are notoriously ill-conditioned: their condition number grows
as the distance between collocation points shrinks and as the lengthscales
grow, a trade-off between approximation power and numerical stability that is
intrinsic to kernel interpolation.
The constrained
Gram matrix is worse in this respect than an ordinary kernel matrix, because it
stacks blocks built from $k$, $\mathcal L k$ and $\mathcal L\mathcal L k$, whose
magnitudes differ by several orders: every application of $\mathcal L$
introduces factors of order $\eta^{-2}$, so the diagonal of the doubly-operated
block scales roughly as $a^{2}\eta^{-4}$ against $a^{2}$ for the boundary block.

We regularise the system by adding a constant $\sigma^*$ to the diagonal:
\begin{equation}\label{eq:Creg}
  C_{\mathrm{reg}}=C+\sigma^{*2}\,\mathbf I_{n_I+n_B}
\end{equation}
In the Gaussian process model, \eqref{eq:Creg} corresponds to independent Gaussian
observation noise on the two families of constraints \citep[Chap.~2]{rasmussen2006gaussian}.
If we concatenate the boundary conditions and interior source terms into an observation vector perturbed by independent Gaussian noise $\boldsymbol{\epsilon} \sim \mathcal{N}(\mathbf{0}, \sigma^{*2} \mathbf I_{n_I+n_B})$:
\begin{equation}
\tilde{\boldsymbol u} = \begin{bmatrix} h(\bar{\boldsymbol X}) \\ z(\mathring{\boldsymbol X}) \end{bmatrix} + \boldsymbol{\epsilon}.
\end{equation}
By the independence of $\boldsymbol{\epsilon}$, the covariance matrix of the observations simplifies to the regularised Gram matrix:
\begin{equation}\label{eq:C_reg}
C_{\mathrm{reg}} = \operatorname{Cov}(\tilde{\boldsymbol u}, \tilde{\boldsymbol u}) = C(\mathring{\boldsymbol X}, \bar{\boldsymbol X}) + \sigma^{*2} \mathbf{I}.
\end{equation}

The regularisation constant is not free. It perturbs the constrained regression, and bounds from below the precision attainable by the stopping criterion of
Section~\ref{sec:stopping}; $\sigma^*$ should accordingly be chosen as the smallest
value for which the Gram matrix is stable.

Linear systems are solved through the Cholesky factorisation
$C_{\mathrm{reg}}=LL^{\top}$ rather than by forming $C_{\mathrm{reg}}^{-1}$. The
factorisation is backward stable, so the computed weights are the exact
solution of a system perturbed at the level of machine precision, and the
forward error is of order $\operatorname{cond}(C_{\mathrm{reg}})$ times machine precision \citep[Chap.~10]{higham2002accuracy}.
Crucially for the fixed-point scheme \eqref{eq:iteration}, the Gram matrix does
not depend on the source $z$: only the right-hand side changes from one
iteration to the next. The factorisation, which costs $O((n_I+n_B)^{3})$, is
therefore computed once, and every subsequent iteration costs $O((n_I+n_B)^{2})$
for the two triangular solves, plus the evaluation of the jump term.

The full pricing algorithm is therefore given in Algorithm~\ref{alg:cgpr}.

\begin{algorithm}[H]
\caption{Computation of the price $\widehat V^{(n)}(x)$ of a derivative with boundary function $h$}
\label{alg:cgpr}
\begin{enumerate}
    \item Sample $n_I$ points $\mathring{\boldsymbol{x}}_i=(\mathring t_i,\mathring{\boldsymbol{y}}_i)\in\mathring{\mathcal X}$ and $n_B$ points $\bar {\boldsymbol x}_i\in\partial\mathcal X$.
    \item Set $h_i=h(\bar {\boldsymbol x}_i)$ and $\boldsymbol z^{(0)}=\boldsymbol 0$.
    \item Compute the Gram matrix
    \begin{equation*}
        C=\sigma^{*2}I_{n_I+n_B}+
        \begin{pmatrix}
        k(\bar X,\bar X) & \mathcal L_{\mathring x}k(\mathring{\boldsymbol{X}},\bar{\boldsymbol{X}})^\top\\
        \mathcal L_{\mathring x}k(\mathring{\boldsymbol{X}},\bar{\boldsymbol{X}}) & \mathcal L_{\mathring x}\mathcal L_{\tilde x}k(\mathring{\boldsymbol{X}},\mathring{\boldsymbol{X}})
        \end{pmatrix}
    \end{equation*}
    and its Cholesky factorisation $C = LL^\top$.\\
    Calculate the vectors $\boldsymbol\beta(\mathring{\boldsymbol x}_i) = \begin{pmatrix} k(\bar{\boldsymbol X}, \mathring{\boldsymbol x}_i) \\ \mathcal L_{\mathring{\boldsymbol x}} k(\mathring{\boldsymbol X}, \mathring{\boldsymbol x}_i) \end{pmatrix}$ for $i = 1, \dots, n_I$.

    Calculate $\mathbb E_{\xi}\bigl[\boldsymbol\beta(\mathring t_i, \mathring{\boldsymbol y}_i + M^{(v)}\xi)^\top\bigr]$ for $v = 1, \dots, p-1$ and $i = 1, \dots, n_I$.

    Calculate $m(\mathring{\boldsymbol x}_i)$, $\mathcal L m(\mathring{\boldsymbol x}_i)$ and $\mathcal J(m)(\mathring{\boldsymbol x}_i)$ for $i = 1, \dots, n_I$, and $m(\bar{\boldsymbol x}_i)$ for $i = 1, \dots, n_B$.
    \item Loop on $n=1,\dots,N(\varepsilon)$, where
    $N(\varepsilon)=\min\bigl\{n\ge1:\ c_T U_{n-1}\Lambda\|h\|_\infty\le\varepsilon\bigr\}$:
    \begin{enumerate}
        \item Solve
        $$LL^\top\boldsymbol\alpha^{(n)}=\begin{pmatrix}\boldsymbol h-m(\bar{\boldsymbol{X}} )\\ \boldsymbol{z}^{(n-1)}-\mathcal Lm(\mathring {\boldsymbol{X}})\end{pmatrix}$$
        and evaluate, for $i=1,\dots,n_I$,
        $$\widehat V^{(n)}(\mathring {\boldsymbol{x}}_i)=m(\mathring {\boldsymbol{x}}_i)+\boldsymbol\beta(\mathring {\boldsymbol{x}}_i)^\top\boldsymbol\alpha^{(n)}.$$
        \item Calculate the vector $\boldsymbol{\mathcal J}^{(n)}$ of
        $\mathcal J\bigl(\widehat V^{(n)}\bigr)(\mathring {\boldsymbol{x}}_i)=\mathcal J(m)(\mathring {\boldsymbol{x}}_i)+\mathcal J(\boldsymbol\beta)(\mathring {\boldsymbol{x}}_i)^\top\boldsymbol\alpha^{(n)}$, $i=1,\dots,n_I$.
        \item Update $\boldsymbol z^{(n)}=-\boldsymbol{\mathcal J}^{(n)}$.
    \end{enumerate}
    \item Return $\widehat V^{(N(\varepsilon))}(x)=m(x)+\boldsymbol\beta(x)^\top\boldsymbol\alpha^{(N(\varepsilon))}$.
\end{enumerate}
\end{algorithm}

\section{Kernels estimation}
\label{sec:kernels}
The regression of Section~\ref{sec:cgpr} requires the kernel $k$ together with
its differential forms $\mathcal L_{\tilde x}k$ and $\mathcal L_x\mathcal L_{\tilde x}k$.
The subscript indicates the argument on which $\mathcal L$ acts. Here, we denote by $A$ the instantaneous covariance of the
state.

The kernel must
therefore be at least twice differentiable in each argument. Rather than
differentiating each kernel separately, we derive these images once for the
whole class of radial kernels, and then specialise them to the Matérn~5/2 and
Gaussian kernels \citep[Chap.~4]{rasmussen2006gaussian}.

\subsection{Radial kernels}
For $x=(t,y)$ and $\tilde x=(\tilde t,\tilde y)$ in
$\mathcal X$, define the scaled distance function
\begin{equation}\label{eq:rho}
  \rho(x,\tilde x)=\Bigl(\frac{(t-\tilde t)^2}{\eta_t^2}
  +\sum_{i=1}^{p-1}\frac{(y_i-\tilde y_i)^2}{\eta_i^2}\Bigr)^{1/2}.
\end{equation}
In this paper, we work with radial kernels. Such kernels have the general form $k(x,\tilde x)=a^2\chi\bigl(\rho(x,\tilde x)\bigr)$ for a
scalar profile $\chi:\mathbb R^+\to\mathbb R$. The amplitude $a^2$ fixes the
scale of the functions and the lengthscales $\eta_t,\eta_1,\dots,\eta_{p-1}$ fix
how fast correlations decay in each direction.

All derivatives of a radial kernel follow from one rule: differentiating a
function of $\rho$ with respect to the state produces the derivative of the
profile divided by $\rho$, times a difference vector (Section~S1 of the online supplementary material).
This suggests the profiles
\begin{equation}\label{eq:phi-rec}
  \phi_1(\rho)=\frac{a^2\chi'(\rho)}{\rho},\qquad
  \phi_{n+1}(\rho)=\frac{\phi_n'(\rho)}{\rho},\quad n\ge1,
\end{equation}
and the scaled differences and lengthscale matrix
\begin{equation}\label{eq:notation-d}
  d_t=\frac{t-\tilde t}{\eta_t^2},\qquad
  d_y=D_\eta(y-\tilde y),\qquad
  D_\eta=\operatorname{diag}\bigl(\eta_1^{-2},\dots,\eta_{p-1}^{-2}\bigr).
\end{equation}
Coefficients without a tilde are evaluated at $x$, those with a tilde at
$\tilde x$: $\mu=\mu(x)$, $\tilde\mu=\mu(\tilde x)$,
and similarly for $A$ and $r$.

\begin{proposition}[Single application]\label{prop:Lk}
For a radial kernel,
\begin{equation}\label{eq:F}
  \mathcal L_{\tilde x}k(x,\tilde x):= F(x,\tilde x)
  =-\tilde r\,k-\phi_1\,\theta+\tfrac12\phi_2\,\Omega+\tfrac12\phi_1\,\tilde\tau_A ,
\end{equation}
where, writing $\phi_n=\phi_n(\rho)$,
\begin{equation*}
  \theta=d_t+\tilde\mu^\top d_y,\qquad
  \Omega=d_y^\top\tilde A\,d_y,\qquad
  \tilde\tau_A=\operatorname{tr}\bigl(\tilde A D_\eta\bigr).
\end{equation*}
\end{proposition}

Applying $\mathcal L$ a second time, now
to the first argument, differentiates $F$; since $\theta$ and $\Omega$ depend on
$x$ through $d_t$ and $d_y$, the same rule raises the order of each
profile by one. This introduces the two scalar functions
\begin{equation}\label{eq:SR}
\begin{aligned}
  F_1&=-\tilde r\,\phi_1-\phi_2\,\theta+\tfrac12\phi_3\,\Omega+\tfrac12\phi_2\,\tilde\tau_A,\\
  F_2&=-\tilde r\,\phi_2-\phi_3\,\theta+\tfrac12\phi_4\,\Omega+\tfrac12\phi_3\,\tilde\tau_A,
\end{aligned}
\end{equation}
which are \eqref{eq:F} with every profile shifted by one and two orders.

\begin{proposition}[Double application]\label{prop:LLk}
For a radial kernel,
\begin{equation}\label{eq:LLk}
\begin{aligned}
  \mathcal L_x\mathcal L_{\tilde x}k
  ={}&-r\,F+F_1\Bigl(d_t+\mu^\top d_y+\tfrac12\tau_A\Bigr)
     +\tfrac12F_2\,d_y^\top A\,d_y
     -\frac{\phi_1}{\eta_t^2}\\
  &-\phi_1\,\mu^\top D_\eta\tilde\mu
   +\phi_2\,\mu^\top D_\eta\tilde A\,d_y
   -\phi_2\,d_y^\top A D_\eta\tilde\mu\\
  &+\phi_3\,d_y^\top A D_\eta\tilde A\,d_y
   +\tfrac12\phi_2\operatorname{tr}\bigl(A D_\eta\tilde A D_\eta\bigr),
\end{aligned}
\end{equation}
with $\tau_A=\operatorname{tr}(AD_\eta)$.
\end{proposition}

The proofs are given in Section~S1 of the online supplementary material. The formulas hold for any
profile, and they involve the profiles only through $\phi_1,\dots,\phi_4$: once
these are known, the Gram matrix of Section~\ref{sec:cgpr} follows without further
differentiation. Each entry costs a fixed number of products of vectors and
matrices of size $p-1$.

\begin{remark}[Coincident points]\label{rem:diag}
On the diagonal of the Gram matrix, $x=\tilde x$, so $\rho=0$ and
$d_t=0$, $d_y=0$. Every term of \eqref{eq:LLk} that contains
$d_t$ or $d_y$ vanishes, provided the profiles grow at most like
$\phi_3=O(\rho^{-1})$ and $\phi_4=O(\rho^{-3})$, and
\begin{equation}\label{eq:LLk-diag}
\begin{aligned}
  \mathcal L_x\mathcal L_{\tilde x}k\big|_{x=\tilde x}
  ={}&-r F^{\circ}+\tfrac12F_1^{\circ}\tau_A-\frac{\phi_1(0)}{\eta_t^2}
     -\phi_1(0)\,\mu^\top D_\eta\mu\\
  &+\tfrac12\phi_2(0)\operatorname{tr}\bigl(AD_\eta AD_\eta\bigr),
\end{aligned}
\end{equation}
with $F^{\circ}=-r\,a^2\chi(0)+\tfrac12\phi_1(0)\tau_A$ and
$F_1^{\circ}=-r\,\phi_1(0)+\tfrac12\phi_2(0)\tau_A$. This limit must be used explicitly
when $\phi_3$ or $\phi_4$ is singular at the origin, since a direct evaluation of
\eqref{eq:LLk} would then produce $0\cdot\infty$.
\end{remark}

\subsection{Matérn 5/2 kernel}\label{sec:maternkernel}

The Matérn~5/2 kernel,
\begin{equation}\label{eq:matern}
  k=a^2\Bigl(1+\sqrt5\rho+\tfrac53\rho^2\Bigr)e^{-\sqrt5\rho},
\end{equation}
generates functions that are twice, but not three times, mean-square
differentiable, which allows the regression to follow the limited smoothness of
option prices near maturity. Applying \eqref{eq:phi-rec} gives
\begin{equation}\label{eq:matern-phi}
\begin{aligned}
  \phi_1&=-\tfrac53a^2\bigl(1+\sqrt5\rho\bigr)e^{-\sqrt5\rho}, &
  \phi_2&=\tfrac{25}{3}a^2e^{-\sqrt5\rho},\\
  \phi_3&=-\tfrac{25\sqrt5}{3}a^2\,\frac{e^{-\sqrt5\rho}}{\rho}, &
  \phi_4&=\tfrac{25}{3}a^2\Bigl(\frac{5}{\rho^2}+\frac{\sqrt5}{\rho^3}\Bigr)e^{-\sqrt5\rho}.
\end{aligned}
\end{equation}
The first two profiles are bounded, with $\phi_1(0)=-\tfrac53a^2$ and
$\phi_2(0)=\tfrac{25}{3}a^2$, while $\phi_3$ and $\phi_4$ are singular at the
origin with exactly the rates of Remark~\ref{rem:diag}. Off the diagonal they are
evaluated as they stand; on the diagonal \eqref{eq:LLk-diag} is used. The double
application \eqref{eq:LLk} thus exhausts the smoothness of the Matérn~5/2
kernel: $\mathcal L_x\mathcal L_{\tilde x}k$ is continuous, but no further
derivative exists at $\rho=0$.

\subsection{Gaussian kernel}\label{sec:gausskernel}

The Gaussian kernel,
\begin{equation}\label{eq:gauss}
  k=a^2e^{-\rho^2/2},
\end{equation}
generates infinitely differentiable functions. Since $\chi'(\rho)=-\rho\,\chi(\rho)$,
each step of \eqref{eq:phi-rec} only changes the sign:
\begin{equation}\label{eq:gauss-phi}
  \phi_n=(-1)^n\,k,\qquad n\ge1 .
\end{equation}
All profiles are bounded, \eqref{eq:LLk} can be evaluated everywhere including
the diagonal, and \eqref{eq:F} reduces to
$F=k\bigl(\theta-\tilde r+\tfrac12\Omega-\tfrac12\tilde\tau_A\bigr)$. The price of
this smoothness is conditioning: for long lengthscales the Gram matrix of the
Gaussian kernel is much closer to singular than that of the Matérn kernel.

\subsection{Kernel parameters fitting}
\label{sec:hyper}
Prior to testing the method on numerical experiments for derivatives pricing, the model must be properly calibrated. To estimate the parameters $a^{2}$ and $\boldsymbol\eta=(\eta_{t},\eta_{1},\dots,\eta_{p-1})$, Gaussian processes usually rely on Maximum Likelihood Estimation \citep[see][Chap.~5]{rasmussen2006gaussian}. However, for CGPR, the likelihood optimisation is very unstable. That's why we opt for a grid search that minimises the mean square PIDE residual:
\begin{equation}\label{eq:R-eta}
  \mathcal R(a, \boldsymbol\eta)
  :=\frac{1}{n_{I,\mathrm{val}}}\sum_{\boldsymbol x\in\mathring{\boldsymbol X}_{\mathrm{val}}}
     \Bigl(\mathcal L\widehat V_{a, \boldsymbol\eta}(\boldsymbol x)
       +\mathcal J\bigl(\widehat V_{a, \boldsymbol\eta}\bigr)(\boldsymbol x)\Bigr)^{2}
  +\frac{w}{n_{B,\mathrm{val}}}\sum_{\bar{\boldsymbol x}\in\bar{\boldsymbol X}_{\mathrm{val}}}
     \Bigl(\widehat V_{a, \boldsymbol\eta}(\bar{\boldsymbol x})-h(\bar{\boldsymbol x})\Bigr)^{2}
\end{equation}
where $\widehat V_{a,\boldsymbol\eta}$ is the surface returned by
Algorithm~\ref{alg:cgpr}, $\mathring{\boldsymbol X}_{\mathrm{val}}$
and $\bar{\boldsymbol X}_{\mathrm{val}}$ being sets of domain points that were not used in the Gram matrix construction, which contain respectively $n_{I,\mathrm{val}}$ and $n_{B,\mathrm{val}}$ points.

\section{Closed-form Greeks}\label{sec:greeks}

A practical advantage of the CGPR is that it returns the price as an explicit,
differentiable function of the state, so that its sensitivities can be obtained
by differentiation rather than by finite differences or additional
simulations. At an arbitrary point $\boldsymbol x=(t,\boldsymbol y)$, the predictive mean is
\begin{equation}\label{eq:post-mean}
  \widehat V(\boldsymbol x)=m(\boldsymbol x)+\beta(\boldsymbol x)^\top\alpha,
  \qquad
  \beta(\boldsymbol x)=\begin{pmatrix}\mathcal L_{\tilde {\boldsymbol x}}k({\boldsymbol x},\tilde {\boldsymbol x})\big|_{\tilde {\boldsymbol x}=\mathring {\boldsymbol X}}\\[2pt]
             k({\boldsymbol x},\bar {\boldsymbol X})\end{pmatrix},
\end{equation}
where $\alpha$ is the weight vector obtained with the Gram matrix $C$ at the
last iterate of Algorithm~\ref{alg:cgpr}, and $\beta(x)$ collects the covariances
between $x$ and the interior points $\mathring X$ and boundary points $\bar X$.
The top block is the function $F$ of \eqref{eq:F}. Since differentiation is linear
and $\alpha$ does not depend on $x$, every derivative of $\widehat V$ passes onto
$m$ and $\beta$.

\begin{lemma}\label{lem:dbeta}
For every state variable $\gamma\in\{t,y_1,\dots,y_{p-1}\}$,
\begin{equation*}
  \frac{\partial\boldsymbol\beta(\boldsymbol x)}{\partial\gamma}
  =\begin{pmatrix}
     \mathcal L_{\tilde{\boldsymbol x}}\bigl(\partial_\gamma k(\boldsymbol x,\tilde{\boldsymbol x})\bigr)\big|_{\tilde{\boldsymbol x}=\mathring{\boldsymbol X}}\\[2pt]
     \partial_\gamma k(\boldsymbol x,\bar{\boldsymbol X})
   \end{pmatrix}
\end{equation*}
and the same holds for second derivatives.
\end{lemma}

\begin{proof}
The coefficients $\tilde r$, $\tilde\mu$ and $\tilde A$ of $\mathcal L_{\tilde x}$ depend
on $\tilde x$ only, and therefore act as constants under $\partial_\gamma$. Since
mixed partial derivatives of $k$ commute, $\partial_\gamma$ and $\mathcal L_{\tilde x}$
can be interchanged.
\end{proof}
Combining \eqref{eq:post-mean} with Lemma~\ref{lem:dbeta} gives every
sensitivity of the price with respect to the state in closed form.
\begin{proposition}\label{prop:greeks}
Let $\mathring{\boldsymbol x}_i$, $i=1,\dots,n_I$, be the interior points and $\bar{\boldsymbol x}_j$,
$j=1,\dots,n_B$, the boundary points, and let $\alpha_i$ and $\alpha_{n_I+j}$ be the
corresponding weights. For all state variables
$\gamma,\gamma_1,\gamma_2\in\{t,y_1,\dots,y_{p-1}\}$,
\begin{align}
  \frac{\partial\widehat V}{\partial\gamma}(\boldsymbol x)
  &=\frac{\partial m}{\partial\gamma}(\boldsymbol x)
   +\sum_{i=1}^{n_I}\alpha_i\,\mathcal L_{\tilde{\boldsymbol x}}
     \Bigl(\frac{\partial k}{\partial\gamma}(\boldsymbol x,\tilde{\boldsymbol x})\Bigr)\Big|_{\tilde{\boldsymbol x}=\mathring{\boldsymbol x}_i}
   +\sum_{j=1}^{n_B}\alpha_{n_I+j}\,\frac{\partial k}{\partial\gamma}(\boldsymbol x,\bar{\boldsymbol x}_j),
   \label{eq:dV}\\
  \frac{\partial^2\widehat V}{\partial\gamma_1\partial\gamma_2}(\boldsymbol x)
  &=\frac{\partial^2 m}{\partial\gamma_1\partial\gamma_2}(\boldsymbol x)
   +\sum_{i=1}^{n_I}\alpha_i\,\mathcal L_{\tilde{\boldsymbol x}}
     \Bigl(\frac{\partial^2k}{\partial\gamma_1\partial\gamma_2}(\boldsymbol x,\tilde{\boldsymbol x})\Bigr)\Big|_{\tilde{\boldsymbol x}=\mathring{\boldsymbol x}_i}
   +\sum_{j=1}^{n_B}\alpha_{n_I+j}\,\frac{\partial^2k}{\partial\gamma_1\partial\gamma_2}(\boldsymbol x,\bar{\boldsymbol x}_j).
   \label{eq:d2V}
\end{align}
In particular, the Theta, Deltas and Gammas of the CGPR price are
$\Theta=\partial_t\widehat V$, $\Delta_v=\partial_{y_v}\widehat V$ and
$\Gamma_{vw}=\partial^2_{y_vy_w}\widehat V$, for $v,w=1,\dots,p-1$.
\end{proposition}

\begin{proof}
Since $\alpha$ does not depend on $x$, differentiating \eqref{eq:post-mean} gives
$\partial_\gamma\widehat V=\partial_\gamma m+\partial_\gamma\beta^\top\alpha$, and
likewise for second derivatives. Lemma~\ref{lem:dbeta} gives the derivatives of
$\beta$.
\end{proof}

\section{Numerical Application}\label{sec:numerics}

We show the generality of the method on a stochastic volatility model
with jumps, the Heston--Kou model. It follows the model of \citet{bates1996jumps}, which adds log-normal jumps to the \citet{heston1993closed} stochastic variance, with the jumps replaced by the double-exponential jumps of \citet{kou2002jump}. We price three European contracts on a
basket of two assets: a call, a put and a double knock-out put. Each asset has
its own stochastic variance, so that the pricing problem is posed in five
dimensions (time, two prices and two variances), with correlated diffusions and
unbounded jumps. In this setting, grid-based PDE solvers become impractical and
no closed-form price is available; the CGPR prices and Greeks are therefore
compared with Monte Carlo estimates.

The CGPR will be benchmarked against Monte Carlo estimates with $4\times10^5$ paths of a full-truncation Euler scheme of $250$ time steps, both for its prices and for the Greeks it outputs. The Monte Carlo simulations are estimated on $18\,000$ combinations of states $(t,\boldsymbol y)$ to perform the analysis on the whole domain.

\subsection{Heston--Kou model}\label{sec:hk}

Under the risk-neutral measure, the price $y_v$ and the instantaneous variance
$\nu_v$ of asset $v\in\{1,2\}$ follow
\begin{equation}\label{eq:hk-sde}
\begin{aligned}
  dy_{v,t}&=(r-\lambda_v\zeta_v)\,y_{v,t}\,dt+\sqrt{\nu_{v,t}}\,y_{v,t}\,dW^{y_v}_t
           +y_{v,t-}\bigl(e^{J^{(v)}}-1\bigr)\,dN^{(v)}_t,\\
  d\nu_{v,t}&=\kappa_v(\bar\nu_v-\nu_{v,t})\,dt+\epsilon_v\sqrt{\nu_{v,t}}\,dW^{\nu_v}_t .
\end{aligned}
\end{equation}
The variance reverts to its long-term mean $\bar\nu_v$ at speed $\kappa_v$, with
volatility of variance $\epsilon_v$. This model reproduces
volatility smile and clustering. The Poisson process
$N^{(v)}$, of intensity $\lambda_v$, adds jumps of log-size $J^{(v)}$, which
generate the heavy tails and the skew of short-term returns. Following
\citet{kou2002jump}, the log-jump has the asymmetric double-exponential density
\begin{equation}\label{eq:kou-density}
  f_{J^{(v)}}(j)=q_v\,\vartheta^+_v\,e^{-\vartheta^+_vj}\,\mathbf 1_{\{j\ge0\}}
  +(1-q_v)\,\vartheta^-_v\,e^{\vartheta^-_vj}\,\mathbf 1_{\{j<0\}},
\end{equation}
where $q_v\in(0,1)$ is the probability of an upward jump and $\vartheta^+_v>1$,
$\vartheta^-_v>0$ are the decay rates of the two tails. For the discounted prices to
be martingales, the drift is compensated by the expected relative jump size,
which a direct integration of \eqref{eq:kou-density} gives in closed form,
\begin{equation}\label{eq:kou-zeta}
  \zeta_v=\mathbb E\bigl[e^{J^{(v)}}-1\bigr]
  =\frac{q_v\vartheta^+_v}{\vartheta^+_v-1}+\frac{(1-q_v)\vartheta^-_v}{\vartheta^-_v+1}-1 .
\end{equation}
The Brownian motions are correlated: $W^{y_v}$ and $W^{\nu_v}$ have correlation
$\varrho_v$, $W^{y_1}$ and $W^{y_2}$ have correlation
$\varrho_{12}$. The parameters used in the numerical analysis are
reported in Table~\ref{tab:hk-params}.

\begin{table}[htbp]
\centering\small
\begin{tabular}{lcccccccc}
\toprule
 & $\kappa_v$ & $\bar\nu_v$ & $\epsilon_v$ & $\varrho_v$ & $\lambda_v$ & $q_v$ & $\vartheta^+_v$ & $\vartheta^-_v$ \\
\midrule
Asset 1 & 1.8 & $0.15^2$ & 0.10 & $-0.3$ & 10 & 0.35 & 30 & 40 \\
Asset 2 & 2.0 & $0.20^2$ & 0.15 & $-0.4$ & 5  & 0.40 & 25 & 35 \\
\bottomrule
\end{tabular}
\caption{Parameters of the Heston--Kou model. The cross-asset correlation is
$\varrho_{12}=0.5$ and the interest rate is $r=0.05$.}
\label{tab:hk-params}
\end{table}

The model enters the framework of Section~\ref{sec:cgpr} with the state
$y=(y_1,\nu_1,y_2,\nu_2)^\top$, so that $p=5$. The drift, the jump vectors and
the relative jump sizes are
\begin{equation}\label{eq:hk-coeffs}
  \mu(t,y)=\begin{pmatrix}(r-\lambda_1\zeta_1)y_1\\ \kappa_1(\bar\nu_1-\nu_1)\\
            (r-\lambda_2\zeta_2)y_2\\ \kappa_2(\bar\nu_2-\nu_2)\end{pmatrix},
  \qquad
  M(t,y)=\operatorname{diag}(y_1,\,0,\,y_2,\,0),\qquad
  \xi^{(v)}=e^{J^{(v)}}-1,
\end{equation}
so that a jump of asset $v$ moves $y_v$ to $y_ve^{J^{(v)}}$ and leaves the
variances continuous. Since the framework is driven by independent Brownian
motions, the correlations are absorbed in the diffusion matrix: if $L_\Sigma$
denotes the lower-triangular Cholesky factor of the correlation matrix of
$(W^{y_1},W^{\nu_1},W^{y_2},W^{\nu_2})$, then $L_\Sigma \boldsymbol W$, with $\boldsymbol W$ a standard
four-dimensional Brownian motion, has the required correlations, and
\begin{equation}\label{eq:hk-sigma}
  \Sigma(t,y)=\operatorname{diag}\bigl(y_1\sqrt{\nu_1},\ \epsilon_1\sqrt{\nu_1},\
  y_2\sqrt{\nu_2},\ \epsilon_2\sqrt{\nu_2}\bigr)\, L_\Sigma .
\end{equation}
For the double knock-out, whose domain is bounded by barriers on the prices, we
work in log-moneyness $x_v=\ln(y_v/K)$, in which the jumps are additive. With
$y=(x_1,\nu_1,x_2,\nu_2)^\top$, It\^o's formula gives
\begin{equation}\label{eq:hk-log}
  \mu(t,y)=\begin{pmatrix}r-\lambda_1\zeta_1-\tfrac12\nu_1\\ \kappa_1(\bar\nu_1-\nu_1)\\
            r-\lambda_2\zeta_2-\tfrac12\nu_2\\ \kappa_2(\bar\nu_2-\nu_2)\end{pmatrix},
  \qquad
  \Sigma(t,y)=\operatorname{diag}\bigl(\sqrt{\nu_1},\ \epsilon_1\sqrt{\nu_1},\
  \sqrt{\nu_2},\ \epsilon_2\sqrt{\nu_2}\bigr)\,L_\Sigma,
\end{equation}
with $M=\operatorname{diag}(1,0,1,0)$ and $\xi^{(v)}=J^{(v)}$.

\subsection{Quadrature and numerical choices}\label{sec:quadrature}

Algorithm~\ref{alg:cgpr} requires the jump expectations
$\mathbb E\bigl[\beta(t, \boldsymbol y+M^{(v)}\xi^{(v)})\bigr]$ of the feature vector at the
interior points. They do not depend on the iterate and are computed once,
before the iterations, in the log-jump $J^{(v)}$. For the call and the put, let
$x^{(v)}(j)$ denote the state $x$ in which $y_v$ is replaced by $y_ve^{j}$.
Splitting \eqref{eq:kou-density} into its two tails and substituting
$u=\vartheta^+_vj$ and $u=-\vartheta^-_vj$ yields two integrals against the weight
$e^{-u}$ on $(0,\infty)$, which we evaluate by Gauss--Laguerre quadrature
\citep{salzer1949table}. With the $N_Q$ roots $u_i$ of the Laguerre polynomial
of degree $N_Q$ and the associated weights $\varpi_i$, which make the rule exact
for polynomials of degree up to $2N_Q-1$,
\begin{equation}\label{eq:gl-vanilla}
  \mathbb E\bigl[\beta\bigl(x^{(v)}(J^{(v)})\bigr)\bigr]
  \approx\sum_{i=1}^{N_Q}\varpi_i\Bigl[q_v\,\beta\bigl(x^{(v)}(u_i/\vartheta^+_v)\bigr)
  +(1-q_v)\,\beta\bigl(x^{(v)}(-u_i/\vartheta^-_v)\bigr)\Bigr].
\end{equation}

For the double knock-out, a jump that crosses a barrier kills the contract, so
that the integrand vanishes outside the corridor and is discontinuous at the
barriers. The contract survives a jump of asset $v$ if and only if
$J^{(v)}\in(\underline j_v,\overline j_v)$, with $\underline j_v=\ln(B_L/K)-x_v$ and $\overline j_v=\ln(B_U/K)-x_v$, and,
the value after a knock-out being zero, the source term of
Section~\ref{sec:fixedpoint} vanishes. The expectation reduces to integrals over
bounded intervals with smooth integrands,
\begin{equation}\label{eq:gl-dko}
\begin{aligned}
  \mathbb E\bigl[\mathbf 1_{\{\underline j_v<J^{(v)}<\overline j_v\}}\,\beta(x+J^{(v)}e_v)\bigr]
  =&q_v\vartheta^+_v\!\int_0^{\overline j_v}\!\beta(x+je_v)\,e^{-\vartheta^+_vj}\,dj\\
  &+(1-q_v)\vartheta^-_v\!\int_{\underline j_v}^{0}\!\beta(x+je_v)\,e^{\vartheta^-_vj}\,dj
\end{aligned}
\end{equation}
where $e_v$ is the unit vector of the coordinate $x_v$. Both integrals are
computed with $N_Q$-point Gauss--Legendre rules mapped onto $[0,\overline j_v]$ and
$[\underline j_v,0]$, which avoids integrating across the discontinuity. Note that, depending on the kernel and the jump model chosen, closed forms of these integrals may
exist, but we apply a quadrature so it works for any combination of them.

\subsection{Common set-up}\label{sec:num-setup}

The three contracts are written on the equally weighted basket
$B=\tfrac12(y_1+y_2)$ with strike $K=100$. Each final fit uses $n_I=2048$ interior
and $n_B=2000$ boundary points.

The chosen prior mean $m(\boldsymbol x)$ is set to the Black--Scholes price of the
contract written on $B$, with an approximation of the basket variance $v(t,y)$ (a first-order Taylor approximation of the instantaneous basket variance):
\begin{equation}
    m(t,\boldsymbol y)=\mathrm{BS}\bigl(\tau,B;K,r,v(t,\boldsymbol y)\bigr)
\end{equation}
This gives a first biased estimate but close enough to the exact solution including stochastic variance and jump terms. For the barrier put, the $\mathrm{BS}$ price includes the approximate probability of the assets staying within the barriers using Black--Scholes formulas.

\subsection{Call}\label{sec:num-call}

We first consider a basket call with maturity $T=2$ and payoff $(B-K)^+$. The
price and variance domains are $[0,y_{\max,v}]$ and $[\nu_{\min,v},\nu_{\max,v}]$ for each asset, where the bounds are chosen following the price domain needed. At maturity the boundary data equal the
payoff; at $y_1=y_2=0$ the call is worthless, and on the upper faces it is deep
in the money and is given its discounted intrinsic value\footnote{This boundary condition is only a mild approximation but greatly improves the stability of the method.} $B-Ke^{-r(T-t)}$. The boundary function $h$ is therefore defined as:

\begin{equation}\label{eq:h-call}
  h(t,y_1,\nu_1,y_2,\nu_2)=
  \begin{cases}
    (B-K)^+, & t=T,\\[2pt]
    \bigl(B-Ke^{-r(T-t)}\bigr)^+, & y_1=y_{\max,1}\ \text{or}\ y_2=y_{\max,2},\\[2pt]
    0, & y_1=y_2=0.
  \end{cases}
\end{equation}
For the call option, we use the Gaussian kernel, whose infinite differentiability also yields smooth Greeks.

Before turning to the prices, we optimise for the kernel parameters through a grid search minimising the PIDE residual $\mathcal R(a, \boldsymbol\eta)$. The obtained values of parameters are reported in Table~\ref{tab:etas_call_veff}.

\begin{table}[htbp]
\centering\small
\caption{Call: kernel and kernel parameters selected by minimising the PIDE residual \eqref{eq:R-eta}. The $\eta$ are the lengthscales of each coordinate and $\hat a^2$ is the prior variance.}
\label{tab:etas_call_veff}
\begin{tabular}{lrrrrrr}
\toprule
Kernel & $\eta_t$ & $\eta_{y_1}$ & $\eta_{\nu_1}$ & $\eta_{y_2}$ & $\eta_{\nu_2}$ & $\hat a^2$ \\
\midrule
Gaussian & 0.903 & 0.086 & 2.205 & 0.105 & 1.594 & 1.74 \\
\bottomrule
\end{tabular}
\end{table}

With these lengthscales, the fitted surface is compared with the Monte Carlo
prices on the evaluation grid. Table~\ref{tab:errors_call_veff} summarises the
errors.

\begin{figure}[htbp]
  \centering
  \includegraphics[width=\textwidth]{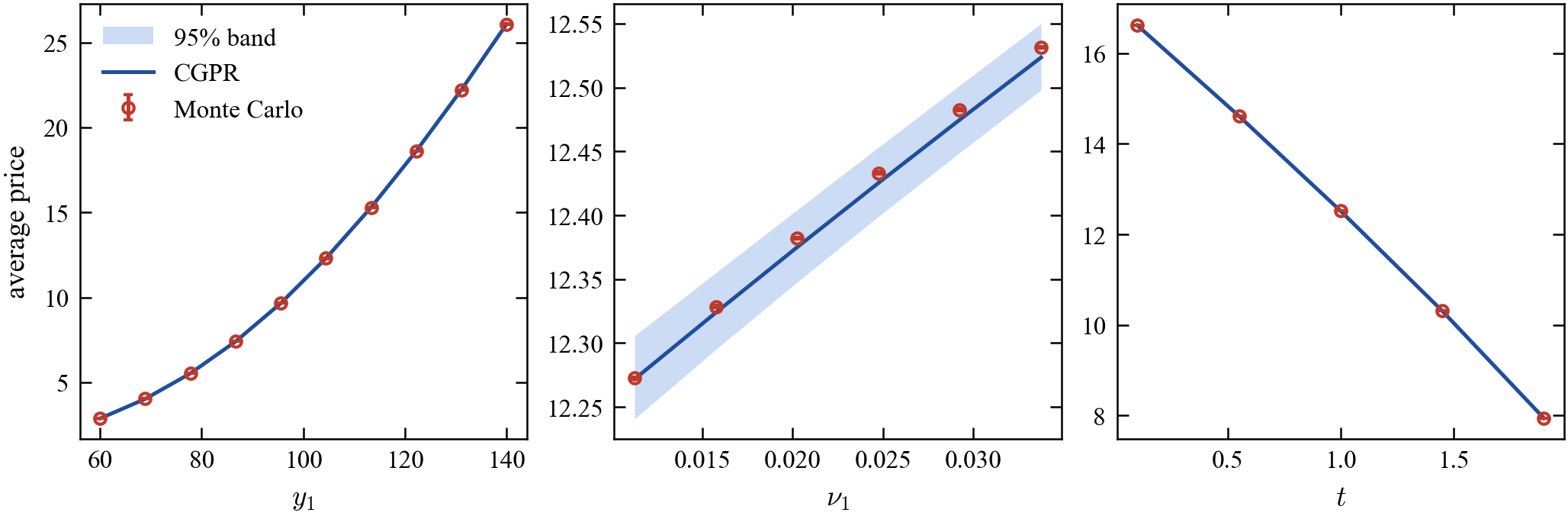}
  \caption{Call: average price as a function of $y_1$, $\nu_1$ and $t$. The line is the
  CGPR posterior mean with its 95\% band; the markers are Monte Carlo prices with
  95\% confidence intervals.}
  \label{fig:profiles-call}
\end{figure}

\begin{figure}[htbp]
  \centering
  \includegraphics[width=0.6\textwidth]{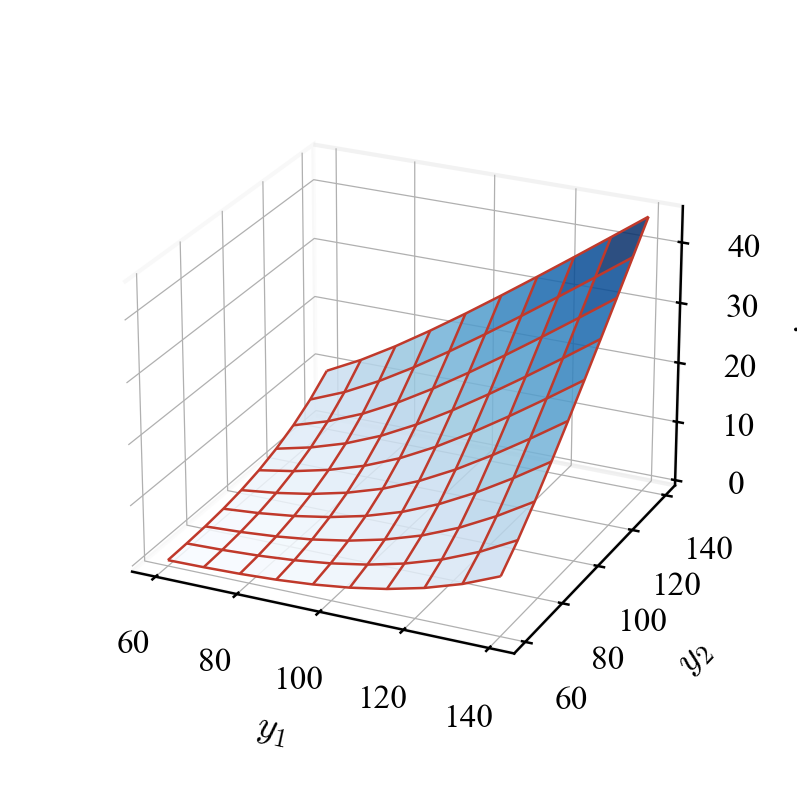}
  \caption{Call: price as a function of $(y_1,y_2)$, averaged over $t$,
  $\nu_1$ and $\nu_2$. The surface is the CGPR posterior mean; the wireframe is
  the Monte Carlo benchmark.}
  \label{fig:surface-call}
\end{figure}

\begin{figure}[htbp]
  \centering
  \includegraphics[width=\textwidth]{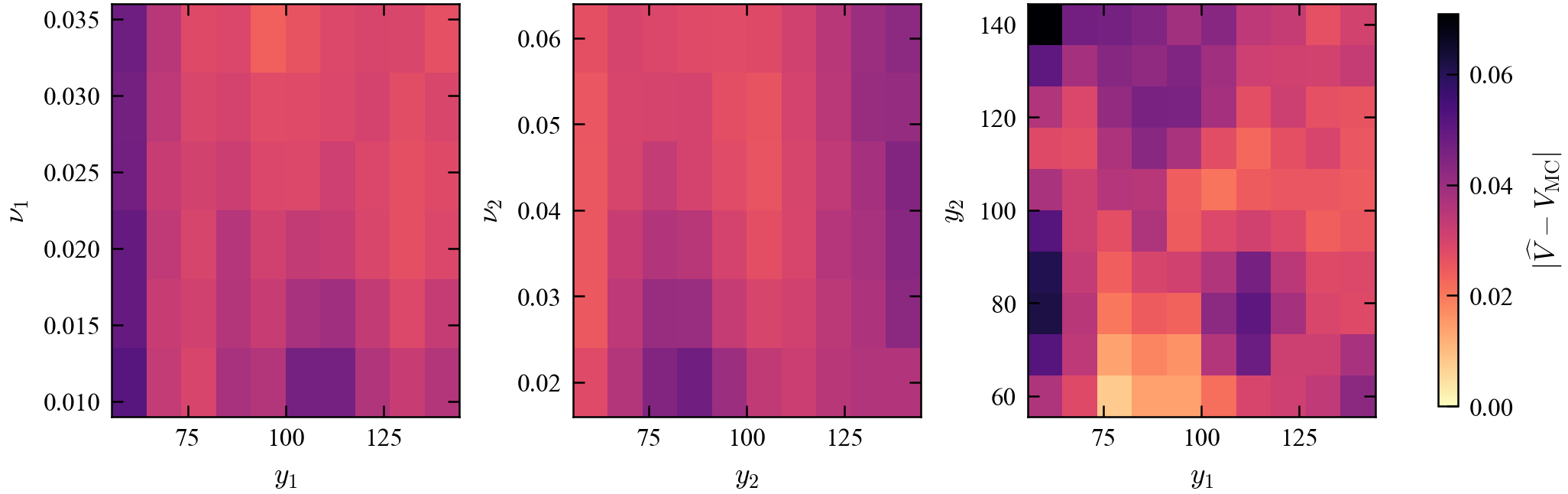}
  \caption{Call: $|\widehat V-V_{\mathrm{MC}}|$ in the planes
  $(y_1,\nu_1)$, $(y_2,\nu_2)$ and $(y_1,y_2)$, averaged over the other
  coordinates.}
  \label{fig:heatmaps-call}
\end{figure}

The quality of the fit is not limited to an average. The profiles of
Figure~\ref{fig:profiles-call}, obtained by averaging the grid over all but one
coordinate, show that the posterior mean follows the benchmark along the price,
the variance and time, and that the
95\% band contains the Monte Carlo prices. Figure~\ref{fig:surface-call} shows
the price surface in the two asset prices, on which the Monte Carlo prices lie.
The error maps of
Figure~\ref{fig:heatmaps-call} show low errors over the domain.

\begin{table}[htbp]
\centering\small
\caption{Call: pricing errors of the CGPR against Monte Carlo (MC) on the test set. Absolute error: $|\widehat V-V_{\mathrm{MC}}|$; relative error: $|\widehat V-V_{\mathrm{MC}}|/V_{\mathrm{MC}}$; MC s.e.: mean standard error of the MC prices.}
\label{tab:errors_call_veff}
\begin{tabular}{rrrrrr}
\toprule
Mean & Mean abs. & Max. abs. & Median rel. & 90\% quantile & MC s.e. \\
MC price & error & error & error (\%) & rel. error (\%) & \\
\midrule
12.405 & 0.0333 & 0.216 & 0.24 & 1.81 & 0.0183 \\
\bottomrule
\end{tabular}
\end{table}

\begin{figure}[htbp]
  \centering
  \includegraphics[width=\textwidth]{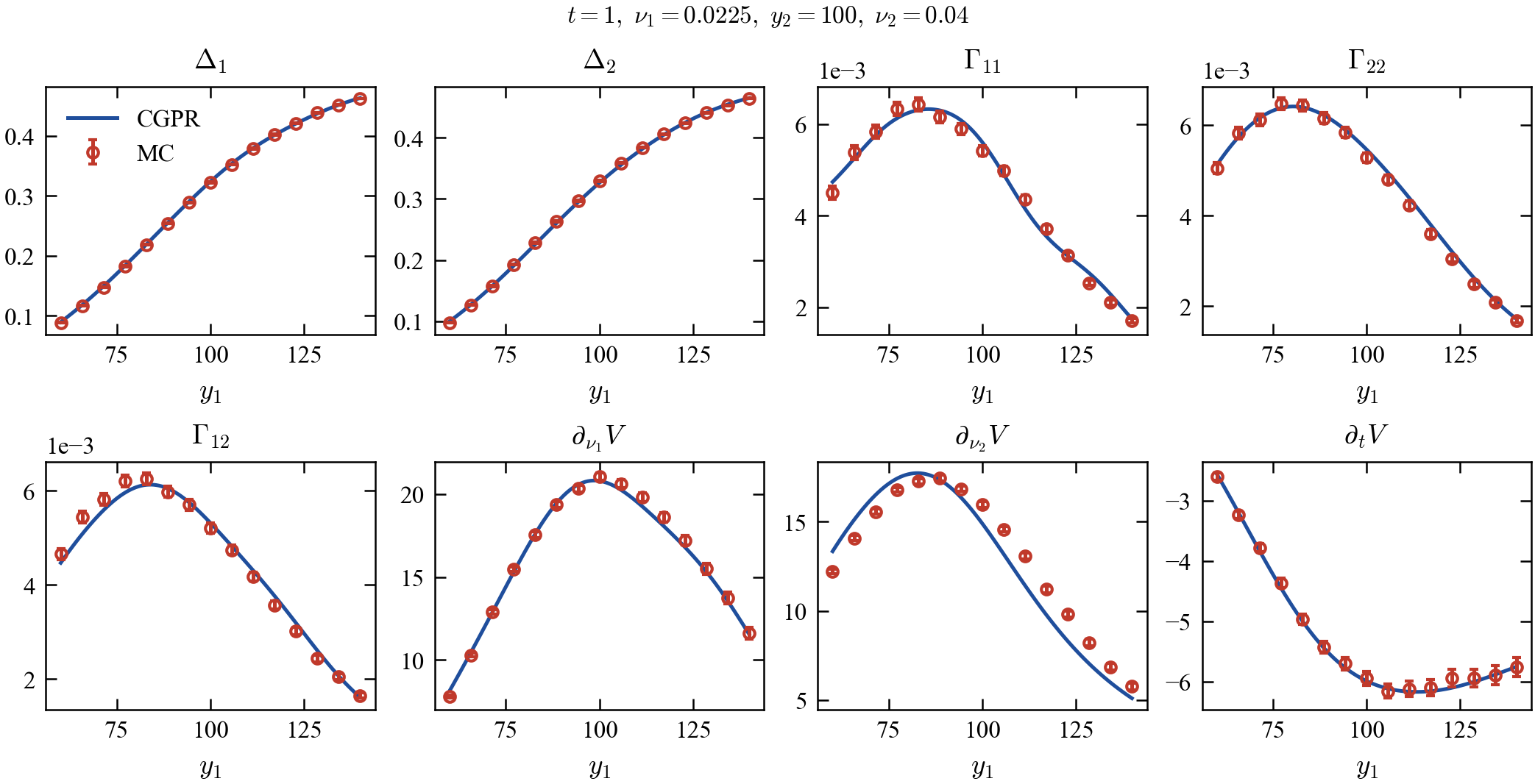}
  \caption{Call: Greeks along $y_1$, the other coordinates being fixed.
  The line is the CGPR; the markers are Monte Carlo estimates with 95\%
  confidence intervals.}
  \label{fig:greeks-call}
\end{figure}

Finally, since the surface is an explicit function of the state, its Greeks
follow directly from Section~\ref{sec:greeks}, without additional fits or
simulations. Table~\ref{tab:greeks_call_veff} compares them with Monte Carlo
estimates on a test set, and Figure~\ref{fig:greeks-call} shows them along
the first price.

\begin{table}[htbp]
\centering\small
\caption{Call: closed-form CGPR sensitivities against Monte Carlo (MC) finite differences on the test set. Rel. error: mean absolute error divided by the mean absolute MC value; MC s.e.: mean standard error of the MC estimates.}
\label{tab:greeks_call_veff}
\begin{tabular}{lrrrr}
\toprule
 & Mean abs. & Mean abs. & Rel. & MC s.e. \\
 & MC value & error & error (\%) & \\
\midrule
$\Delta_1$ & 0.2946 & 0.0022 & 0.75 & 0.000358 \\
$\Delta_2$ & 0.2995 & 0.00134 & 0.45 & 0.000371 \\
$\Gamma_{11}$ & 0.004093 & 0.000225 & 5.51 & 0.0000498 \\
$\Gamma_{22}$ & 0.004072 & 0.000115 & 2.83 & 0.0000496 \\
$\Gamma_{12}$ & 0.003965 & 0.0000983 & 2.48 & 0.0000482 \\
$\partial_{\nu_1}V$ & 13.6 & 1.19 & 8.71 & 0.11 \\
$\partial_{\nu_2}V$ & 10.33 & 0.784 & 7.59 & 0.0804 \\
$\partial_t V$ & 4.85 & 0.0753 & 1.55 & 0.0561 \\
\bottomrule
\end{tabular}
\end{table}

\subsection{Put}\label{sec:num-put}

We then consider a basket put with maturity $T=2$ and payoff $(K-B)^+$. The
price and variance domains are the same as for the basket call. At maturity the boundary data equal the
payoff; at $y_1$ or $y_2=0$ the put is deep in the money and we set its value to its discounted payoff $Ke^{-r(T-t)} - B$, and on the upper faces it is deep out of the money and is given a $0$ value\footnote{Those boundary conditions are only a mild approximation but greatly improve the stability of the method.}. The boundary function $h$ is therefore defined as:
\begin{equation}\label{eq:h-put}
  h(t,y_1,\nu_1,y_2,\nu_2)=
  \begin{cases}
    (K-B)^+, & t=T,\\[2pt]
    Ke^{-r(T-t)}-B, & y_1=0\ \text{or}\ y_2=0,\\[2pt]
    0, & y_1=y_{\max,1}\ \text{or}\ y_2=y_{\max,2}.
  \end{cases}
\end{equation}
The put price has limited smoothness close to maturity at the money, so we use
the Matérn~5/2 kernel.

The lengthscales are selected as for the call.
The optimal parameters are reported in Table~\ref{tab:etas_put_veff}.

\begin{table}[htbp]
\centering\small
\caption{Put: kernel and kernel parameters selected by minimising the PIDE residual \eqref{eq:R-eta}. The $\eta$ are the lengthscales of each coordinate and $\hat a^2$ is the prior variance.}
\label{tab:etas_put_veff}
\begin{tabular}{lrrrrrr}
\toprule
Kernel & $\eta_t$ & $\eta_{y_1}$ & $\eta_{\nu_1}$ & $\eta_{y_2}$ & $\eta_{\nu_2}$ & $\hat a^2$ \\
\midrule
Mat\'ern 5/2 & 5.837 & 0.554 & 2.336 & 0.618 & 1.363 & 30.9 \\
\bottomrule
\end{tabular}
\end{table}

\begin{figure}[htbp]
  \centering
  \includegraphics[width=\textwidth]{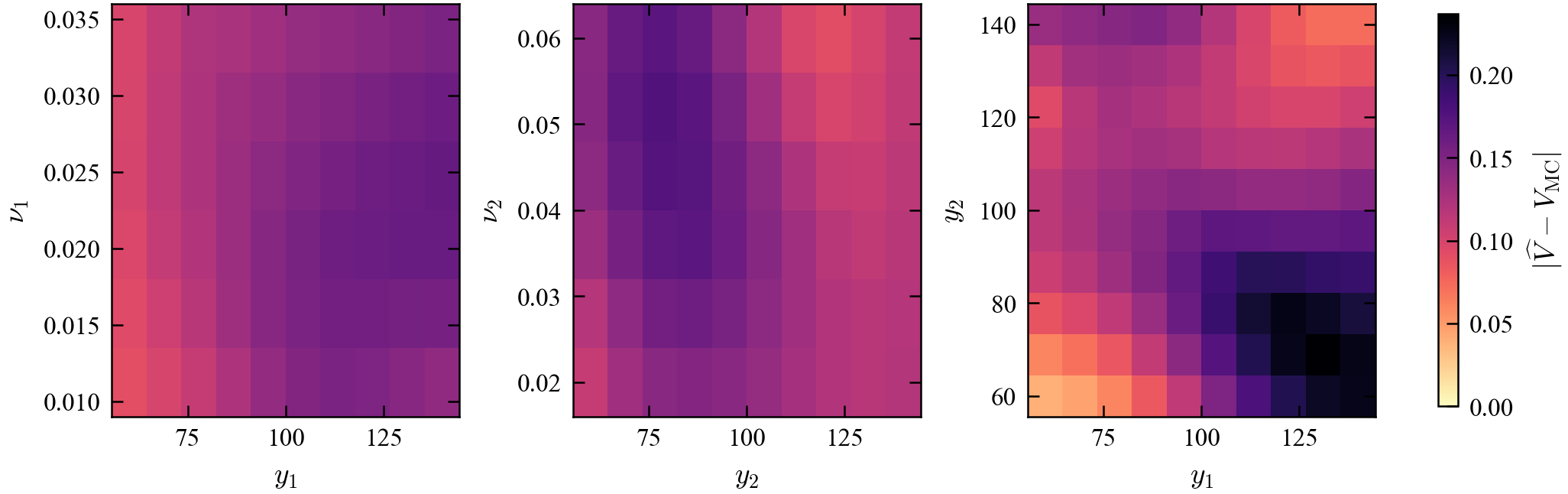}
  \caption{Put: $|\widehat V-V_{\mathrm{MC}}|$ in the planes $(y_1,\nu_1)$,
  $(y_2,\nu_2)$ and $(y_1,y_2)$.}
  \label{fig:heatmaps-put}
\end{figure}

With these lengthscales, Table~\ref{tab:errors_put_veff} summarises the errors
with respect to Monte Carlo.
Figure~\ref{fig:heatmaps-put} shows higher errors for combinations of high values of asset $y_1$ and low values of $y_2$. As those assets are similarly modelled, it may be due to the sparsity of the sampled points in this part of the domain, sampling yielding poorer estimates.

\begin{figure}[htbp]
  \centering
  \includegraphics[width=\textwidth]{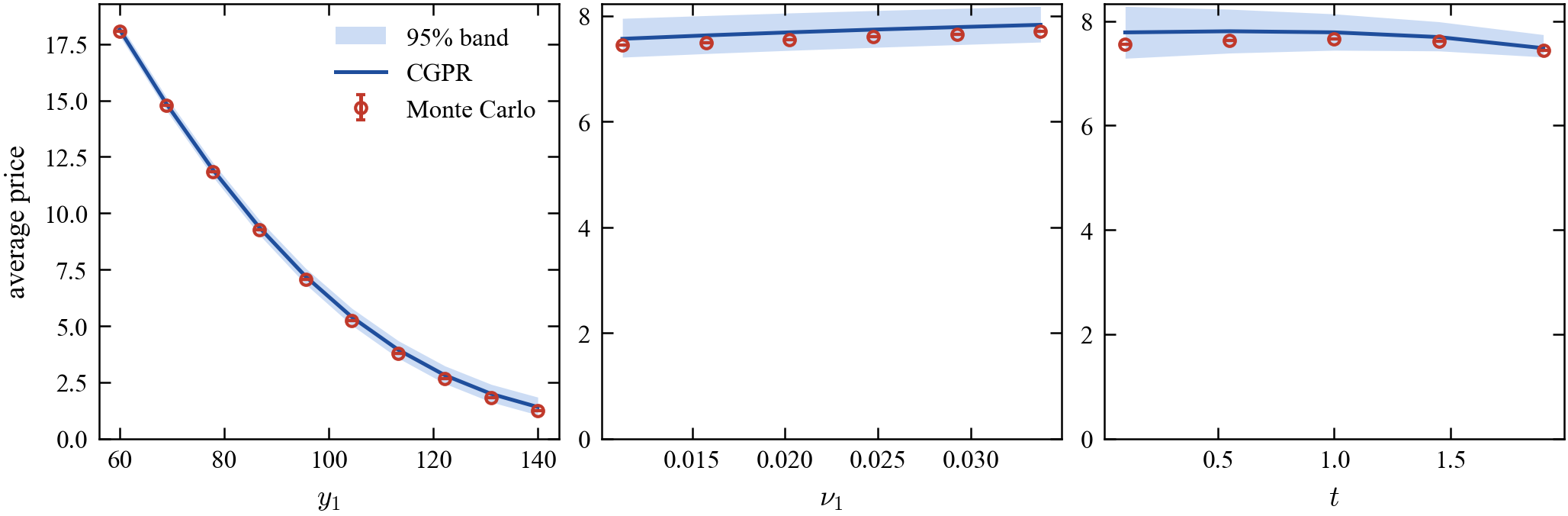}
  \caption{Put: average price along $y_1$, $\nu_1$ and $t$. CGPR with 95\% band; Monte Carlo with 95\% intervals.}
  \label{fig:profiles-put}
\end{figure}

\begin{figure}[htbp]
  \centering
  \includegraphics[width=0.6\textwidth]{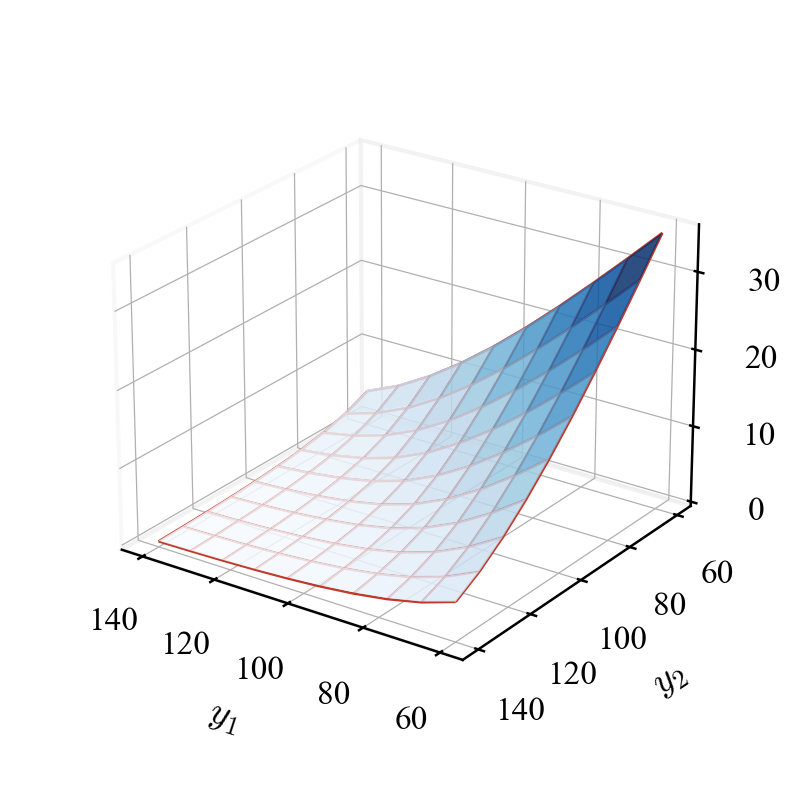}
  \caption{Put: price in $(y_1,y_2)$, averaged over $t$, $\nu_1$, $\nu_2$. Surface:
  CGPR; wireframe: Monte Carlo.}
  \label{fig:surface-put}
\end{figure}

The profiles of Figure~\ref{fig:profiles-put} show that the posterior mean
follows the benchmark along the price, the variance and time, and that the 95\%
band contains the Monte Carlo prices. Figure~\ref{fig:surface-put} shows the
price surface in the two asset prices.

\begin{table}[htbp]
\centering\small
\caption{Put: pricing errors of the CGPR against Monte Carlo (MC) on the test set. Absolute error: $|\widehat V-V_{\mathrm{MC}}|$; relative error: $|\widehat V-V_{\mathrm{MC}}|/V_{\mathrm{MC}}$; MC s.e.: mean standard error of the MC prices.}
\label{tab:errors_put_veff}
\begin{tabular}{rrrrrr}
\toprule
Mean & Mean abs. & Max. abs. & Median rel. & 90\% quantile & MC s.e. \\
MC price & error & error & error (\%) & rel. error (\%) & \\
\midrule
7.576 & 0.1359 & 0.485 & 2.82 & 12.44 & 0.0104 \\
\bottomrule
\end{tabular}
\end{table}

\begin{figure}[htbp]
  \centering
  \includegraphics[width=\textwidth]{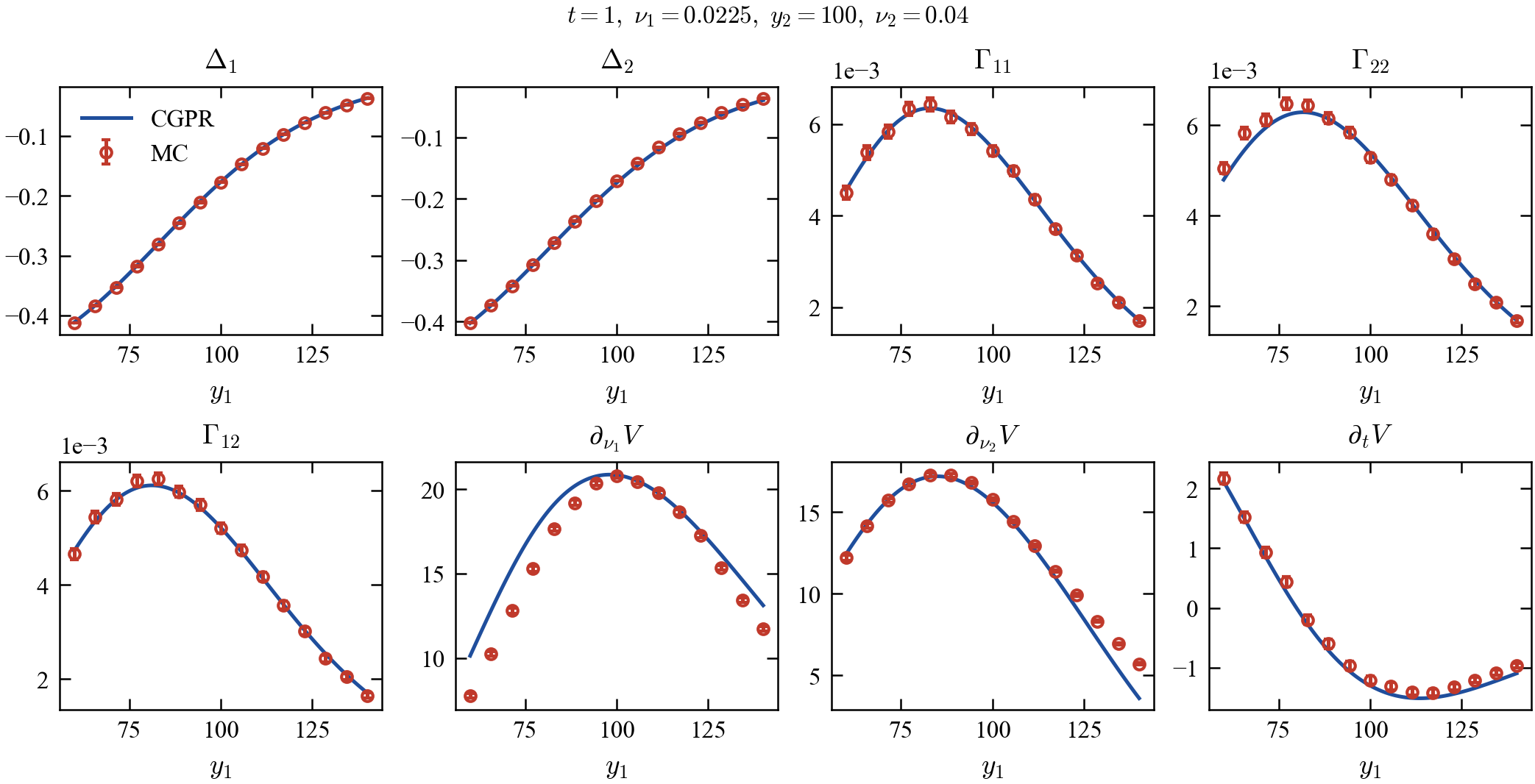}
  \caption{Put: Greeks along $y_1$. CGPR (line) and Monte Carlo with 95\%
  intervals.}
  \label{fig:greeks-put}
\end{figure}

The Greeks are compared with Monte Carlo on the test set in
Table~\ref{tab:greeks_put_veff}, and Figure~\ref{fig:greeks-put} shows them along
the first price.

\begin{table}[htbp]
\centering\small
\caption{Put: closed-form CGPR sensitivities against Monte Carlo (MC) finite differences on the test set. Rel. error: mean absolute error divided by the mean absolute MC value; MC s.e.: mean standard error of the MC estimates.}
\label{tab:greeks_put_veff}
\begin{tabular}{lrrrr}
\toprule
 & Mean abs. & Mean abs. & Rel. & MC s.e. \\
 & MC value & error & error (\%) & \\
\midrule
$\Delta_1$ & 0.2054 & 0.00196 & 0.95 & 0.000269 \\
$\Delta_2$ & 0.2005 & 0.00216 & 1.08 & 0.000262 \\
$\Gamma_{11}$ & 0.004093 & 0.000112 & 2.74 & 0.0000498\ \\
$\Gamma_{22}$ & 0.004072 & 0.000146 & 3.58 & 0.0000496\\
$\Gamma_{12}$ & 0.003965 & 0.0000908 & 2.29 & 0.0000482 \\
$\partial_{\nu_1}V$ & 13.6 & 1.84 & 13.49 & 0.0708 \\
$\partial_{\nu_2}V$ & 10.33 & 1.69 & 16.32 & 0.0491 \\
$\partial_t V$ & 1.42 & 0.108 & 7.62 & 0.0331 \\
\bottomrule
\end{tabular}
\end{table}

\subsection{Barrier Put}\label{sec:num-dko}

Finally, we consider a double knock-out (DKO) put with maturity $T=1$ and payoff
$(K-B)^+$, cancelled as soon as $y_1$ or $y_2$ leaves the corridor
$[B_L,B_U]=[70,150]$. This time, the price domain is the barriers corridor itself, written in log-moneyness, and the variance
domain is as for the call and put. The method is highly practical for barrier options as it does not require approximation on the domain anymore. At maturity, the boundary condition is the payoff $(K-B)^+$ and, on the rest of the domain boundary, the option is deactivated and the value of the option becomes $0$:
\begin{equation}\label{eq:h-dko}
  h(t,y_1,\nu_1,y_2,\nu_2)=
  \begin{cases}
    (K-B)^+, & t=T,\ y_1,y_2\in(B_L,B_U),\\[2pt]
    0, & y_1\in\{B_L,B_U\}\ \text{or}\ y_2\in\{B_L,B_U\}
  \end{cases}
\end{equation}
The price vanishes on the barriers and has steep gradients next to them, so we use the
Matérn~5/2 kernel.

The best kernel parameters for the barrier option are reported in Table~\ref{tab:etas_dko_veff}.

\begin{table}[htbp]
\centering\small
\caption{Double knock-out put: kernel and kernel parameters selected by minimising the PIDE residual \eqref{eq:R-eta}. The $\eta$ are the lengthscales of each coordinate and $\hat a^2$ is the prior variance.}
\label{tab:etas_dko_veff}
\begin{tabular}{lrrrrrr}
\toprule
Kernel & $\eta_t$ & $\eta_{y_1}$ & $\eta_{\nu_1}$ & $\eta_{y_2}$ & $\eta_{\nu_2}$ & $\hat a^2$ \\
\midrule
Mat\'ern 5/2 & 1.651 & 1.057 & 8.383 & 1.406 & 7.522 & 730 \\
\bottomrule
\end{tabular}
\end{table}

\begin{figure}[htbp]
  \centering
  \includegraphics[width=\textwidth]{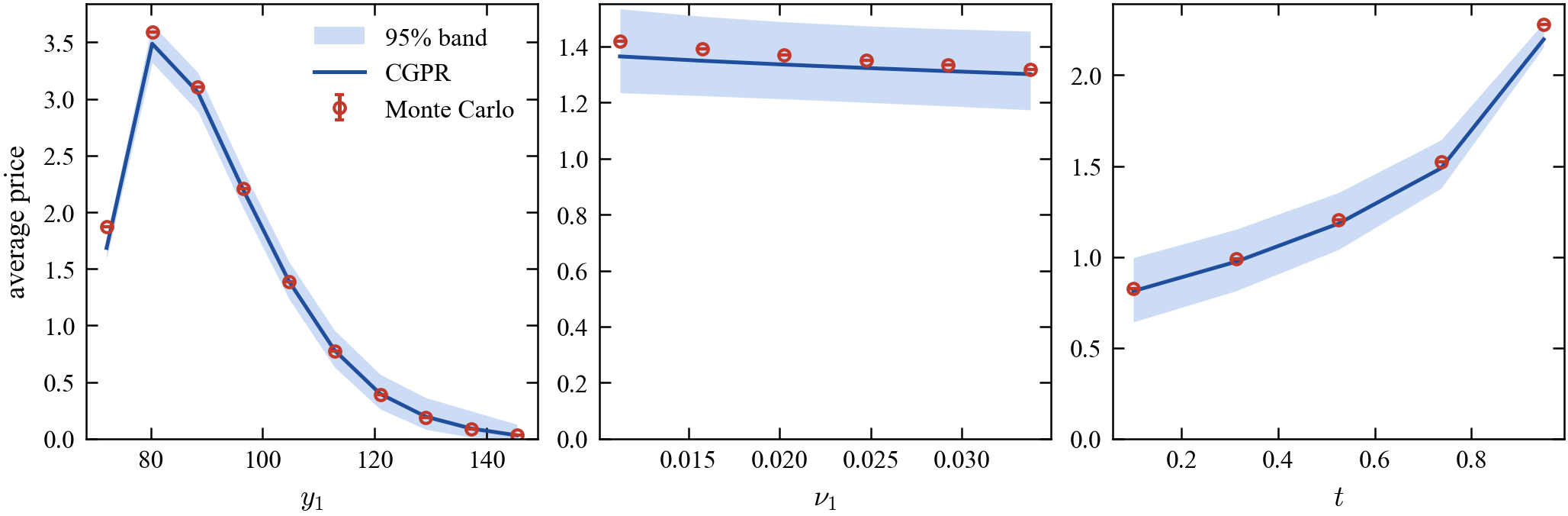}
  \caption{Barrier put: average price along $y_1$, $\nu_1$ and $t$. CGPR with 95\% band; Monte Carlo with 95\% intervals.}
  \label{fig:profiles-dko}
\end{figure}

\begin{figure}[htbp]
  \centering
  \includegraphics[width=0.6\textwidth]{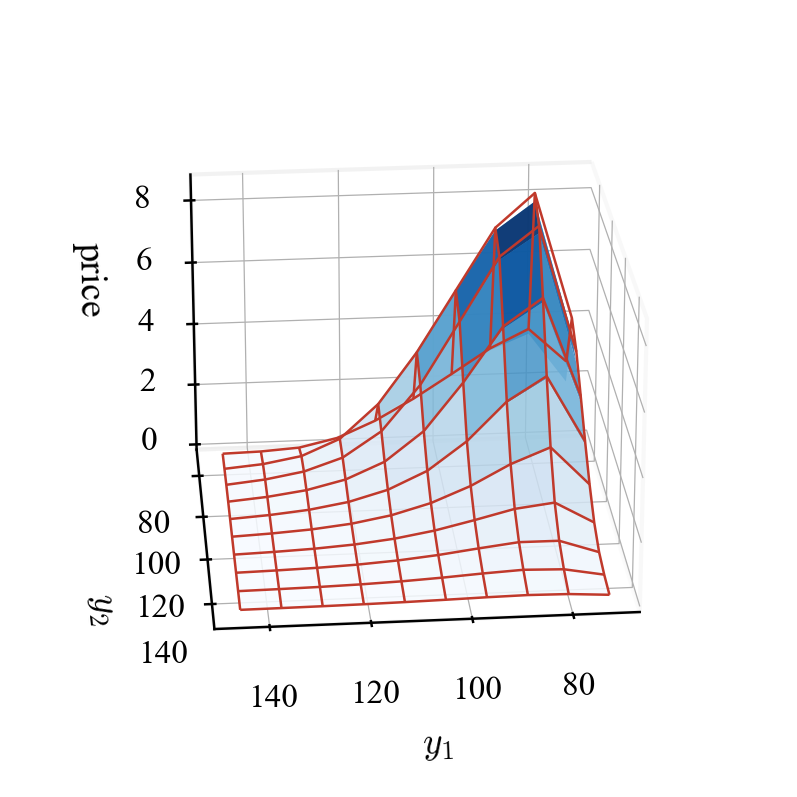}
  \caption{Barrier put: price in $(y_1,y_2)$, averaged over $t$, $\nu_1$, $\nu_2$.
  Surface: CGPR; wireframe: Monte Carlo.}
  \label{fig:surface-dko}
\end{figure}

\begin{figure}[htbp]
  \centering
  \includegraphics[width=\textwidth]{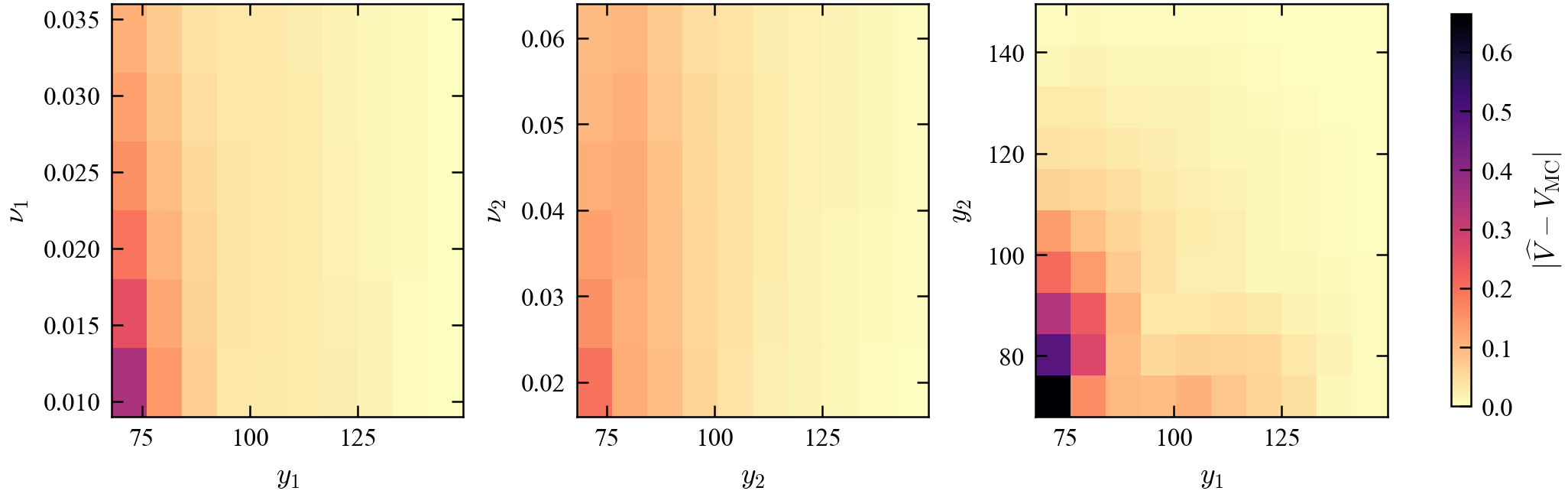}
  \caption{Barrier put: $|\widehat V-V_{\mathrm{MC}}|$ in the planes $(y_1,\nu_1)$,
  $(y_2,\nu_2)$ and $(y_1,y_2)$.}
  \label{fig:heatmaps-dko}
\end{figure}

The profiles of Figure~\ref{fig:profiles-dko} reproduce the shape of the
price between the barriers, and Figure~\ref{fig:surface-dko} shows the price
surface vanishing on the four barriers. The error maps of
Figure~\ref{fig:heatmaps-dko} show that the largest errors happen for combinations of low values of assets and variances.

\begin{table}[htbp]
\centering\small
\caption{Double knock-out put: pricing errors of the CGPR against Monte Carlo (MC) on the test set. Absolute error: $|\widehat V-V_{\mathrm{MC}}|$; relative error: $|\widehat V-V_{\mathrm{MC}}|/V_{\mathrm{MC}}$; MC s.e.: mean standard error of the MC prices.}
\label{tab:errors_dko_veff}
\begin{tabular}{rrrrrr}
\toprule
Mean & Mean abs. & Max. abs. & Median rel. & 90\% quantile & MC s.e. \\
MC price & error & error & error (\%) & rel. error (\%) & \\
\midrule
1.364 & 0.0488 & 4.020 & 2.36 & 6.87 & 0.0036 \\
\bottomrule
\end{tabular}
\end{table}

\begin{figure}[htbp]
  \centering
  \includegraphics[width=\textwidth]{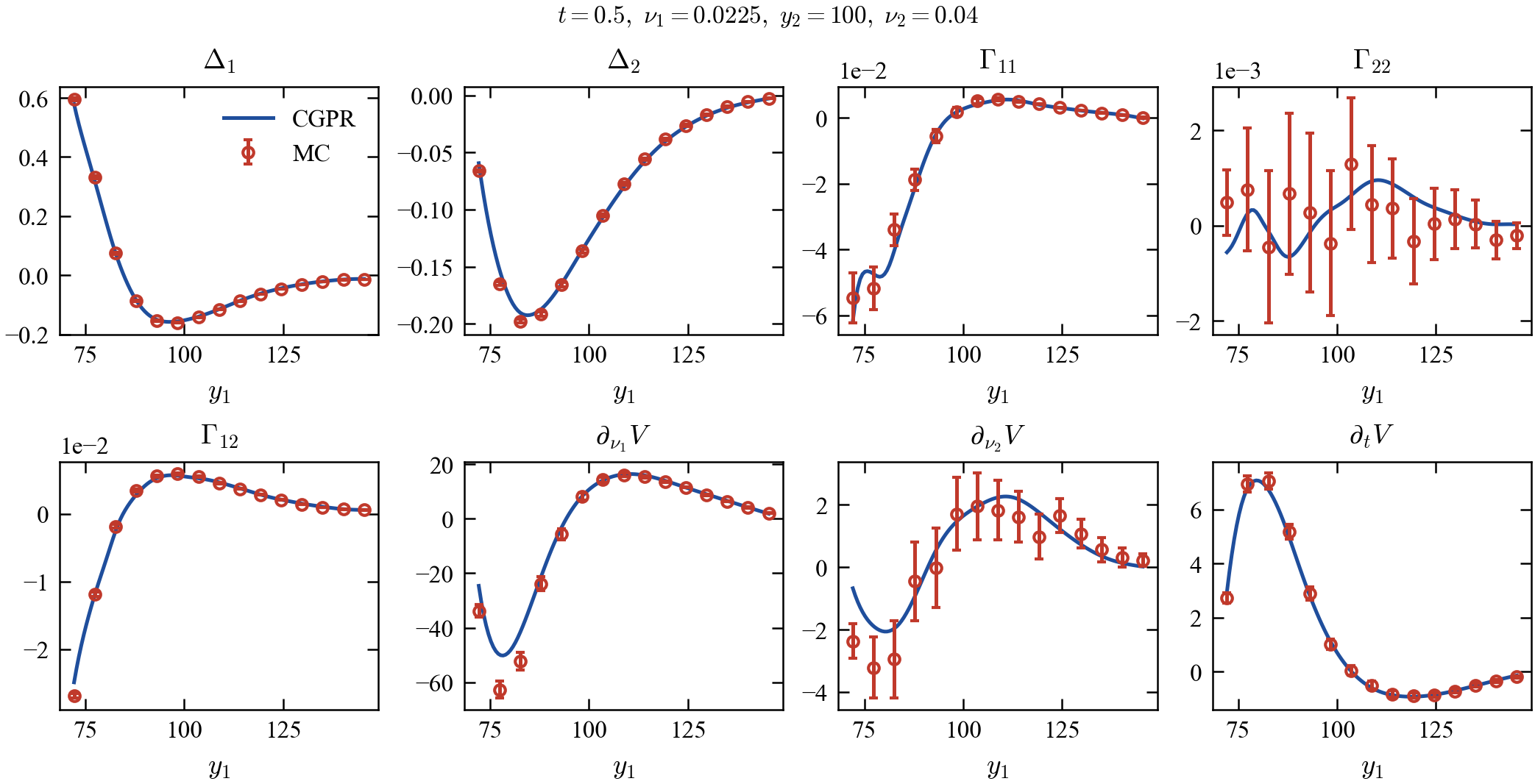}
  \caption{Barrier put: Greeks along $y_1$. CGPR (line) and Monte Carlo with 95\%
  intervals.}
  \label{fig:greeks-dko}
\end{figure}

For the barrier option, the Monte Carlo Greeks are central differences with
common random numbers, and are therefore noisier than for the call and the put.
Table~\ref{tab:greeks_dko_veff} compares them with the CGPR Greeks on the test set, and Figure~\ref{fig:greeks-dko} shows them along the
first price. These figures emphasise the variance in the Greeks for the Monte Carlo method. The smoothness of the CGPR estimates yields closed-form Greeks that are smoother than the ones obtained with MC.

\begin{table}[htbp]
\centering\small
\caption{Double knock-out put: closed-form CGPR sensitivities against Monte Carlo (MC) finite differences on the test set. Rel. error: mean absolute error divided by the mean absolute MC value; MC s.e.: mean standard error of the MC estimates.}
\label{tab:greeks_dko_veff}
\begin{tabular}{lrrrr}
\toprule
 & Mean abs. & Mean abs. & Rel. & MC s.e. \\
 & MC value & error & error (\%) & \\
\midrule
$\Delta_1$ & 0.08028 & 0.00343 & 4.27 & 0.00057 \\
$\Delta_2$ & 0.07381 & 0.00316 & 4.28 & 0.000605 \\
$\Gamma_{11}$ & 0.009597 & 0.00102 & 10.60 & 0.00062 \\
$\Gamma_{22}$ & 0.008222 & 0.000876 & 10.66 & 0.00066 \\
$\Gamma_{12}$ & 0.003981 & 0.000421 & 10.58 & 0.000069 \\
$\partial_{\nu_1}V$ & 10.09 & 1.45 & 14.38 & 0.436 \\
$\partial_{\nu_2}V$ & 8.117 & 0.841 & 10.36 & 0.322 \\
$\partial_t V$ & 2.298 & 0.119 & 5.16 & 0.0608 \\
\bottomrule
\end{tabular}
\end{table}

\section{Conclusion}\label{sec:conclusion}

We have developed a constrained Gaussian process regression for pricing European
derivatives under multivariate jump-diffusions. The prior is conditioned on the
Feynman--Kac PIDE at collocation points, so no exogenous price is needed. The jump
operator is kept out of the covariance and treated as a lagged source, and the
price is the limit of local problems that share one factorised Gram matrix. We
proved that this iteration converges in the supremum norm from any starting
point, that errors in the local solves are amplified by a bounded factor only,
and gave a worst-case number of iterations for a prescribed precision.

The price is returned as a closed-form function of the whole state, and every
state sensitivity follows by differentiating the kernel. On two assets baskets
under the Heston--Kou model, a five-dimensional problem, the algorithm was tried on a call, put and barrier put. Both their prices and sensitivities agreed with the Monte Carlo benchmark. The optimisation of the kernel parameters is the main lever for improvement as it will allow a faster convergence. Natural extensions are regime-switching models, whose
coupling is another non-local term, a boosting scheme fitting successive
regressions to the PIDE residual to reduce the dependence on the prior mean, and
random Fourier features to reduce the cubic cost of the Gram matrix and reach
higher dimensions.

\section*{Acknowledgements}
Computational resources have been provided by the supercomputing facilities of
the Universit\'e catholique de Louvain (CISM/UCL) and the Consortium des
\'Equipements de Calcul Intensif en F\'ed\'eration Wallonie Bruxelles (C\'ECI)
funded by the Fonds de la Recherche Scientifique de Belgique (F.R.S.-FNRS) under
convention 2.5020.11 and by the Walloon Region.

\section*{Author contributions (CRediT)}
\textbf{Arnaud Servais}: Conceptualization, Validation, Writing – original draft, Writing – review \& editing.\\
\textbf{Donatien Hainaut}: Conceptualization, Supervision, Validation, Writing – original draft, Writing – review \& editing.

\section*{Funding}
This work was funded by the UCLouvain - FSR grant for the project “Bayesian regression, generative diffusion and neural estimator for actuarial risks modelling”.

\section*{Disclosure statement}
No potential conﬂict of interest was reported by the authors.

\section*{Declaration of generative AI use}
During the preparation of this work, Claude (Anthropic) was used for
language refinement and as a software coding assistant. The authors reviewed and edited all generated content
and take full responsibility for the content of this article.

\section*{Data availability statement}
Data sharing is not applicable to this article as no new data were created or analyzed in this study.

\bibliographystyle{qf}
\bibliography{references}

\processdelayedfloats
\clearpage
\listoffigures
\end{document}


\title{Supplementary material for\\ `European derivatives pricing via constrained Gaussian process regressions'}
\author{Arnaud Servais and Donatien Hainaut}
\date{}
\maketitle
\section{Derivatives of radial kernels}\label{app:kernels}

This appendix derives the single and double applications of the operator
\begin{equation*}
  \mathcal L_xV=\partial_tV-r(x)V+\mu_y(x)^\top\nabla_yV
  +\tfrac12\operatorname{tr}\bigl(A(x)\,\nabla^2_yV\bigr),
  \qquad A(x)=\Sigma_y(x)\Sigma_y(x)^\top,
\end{equation*}
to a radial kernel $k(x,\tilde x)=k(\rho)$. To keep the formulas short, quantities
evaluated at $x$ are written without a tilde and quantities evaluated at
$\tilde x$ with a tilde:
\begin{equation*}
  \mu=\mu_y(x),\quad \tilde\mu=\mu_y(\tilde x),\quad
  A=A(x),\quad \tilde A=A(\tilde x),\quad
  r=r(x),\quad \tilde r=r(\tilde x).
\end{equation*}
The matrix $A$ is the instantaneous covariance of the state, not to be confused with the Gram matrix; both $A$ and $\tilde A$ are symmetric.

\subsection{Distance and scaled differences}

The anisotropic distance is
\begin{equation*}
  \rho(x,\tilde x)=\Bigl(\frac{(t-\tilde t)^2}{\eta_t^2}
  +\sum_{i=1}^{p-1}\frac{(y_i-\tilde y_i)^2}{\eta_i^2}\Bigr)^{1/2}.
\end{equation*}
We introduce the scaled differences and the matrix of inverse squared
lengthscales,
\begin{equation*}
  d_t=\frac{t-\tilde t}{\eta_t^2},\qquad
  d_y=\Bigl(\frac{y_1-\tilde y_1}{\eta_1^2},\dots,
            \frac{y_{p-1}-\tilde y_{p-1}}{\eta_{p-1}^2}\Bigr)^{\!\top},\qquad
  D_\eta=\operatorname{diag}\Bigl(\frac1{\eta_1^2},\dots,\frac1{\eta_{p-1}^2}\Bigr).
\end{equation*}
Differentiating $\rho^2$ gives $2\rho\,\partial_t\rho=2d_t$ and
$2\rho\,\nabla_y\rho=2d_y$, hence
\begin{align*}
  \partial_t\rho&=\frac{d_t}{\rho}, &
  \nabla_y\rho&=\frac{d_y}{\rho}, &
  \partial_{\tilde t}\rho&=-\frac{d_t}{\rho}, &
  \nabla_{\tilde y}\rho&=-\frac{d_y}{\rho}.
\end{align*}
The scaled differences themselves have the derivatives
\begin{align*}
  \partial_td_t&=\frac1{\eta_t^2}, &
  \nabla_yd_y&=D_\eta, &
  \partial_{\tilde t}d_t&=-\frac1{\eta_t^2}, &
  \nabla_{\tilde y}d_y&=-D_\eta .
\end{align*}

\subsection{The radial rule and the cascade of profiles}

For any differentiable function $g$ of $\rho$ the chain rule gives
\begin{equation*}
  \nabla_yg(\rho)=g'(\rho)\,\nabla_y\rho=\frac{g'(\rho)}{\rho}\,d_y,
  \qquad
  \partial_tg(\rho)=\frac{g'(\rho)}{\rho}\,d_t,
\end{equation*}
and the same with a minus sign for the derivatives in $\tilde x$. A derivative
of a radial function is thus again a radial function, $g'(\rho)/\rho$, times a
scaled difference. This motivates the family of profiles
\begin{equation*}
  \phi_1(\rho)=\frac{k'(\rho)}{\rho},\qquad
  \phi_2(\rho)=\frac{\phi_1'(\rho)}{\rho},\qquad
  \phi_3(\rho)=\frac{\phi_2'(\rho)}{\rho},\qquad
  \phi_4(\rho)=\frac{\phi_3'(\rho)}{\rho},
\end{equation*}
for which the gradients form a cascade:
\begin{align*}
  \nabla_yk&=\phi_1\,d_y, & \nabla_y\phi_1&=\phi_2\,d_y, &
  \nabla_y\phi_2&=\phi_3\,d_y, & \nabla_y\phi_3&=\phi_4\,d_y,
\end{align*}
and likewise $\partial_t\phi_n=\phi_{n+1}d_t$. In what follows $\phi_n$ stands
for $\phi_n(\rho)$.

\subsection{Derivatives of the kernel}

With respect to $\tilde x$, the cascade gives
\begin{equation*}
  \partial_{\tilde t}k=-\phi_1\,d_t,\qquad
  \nabla_{\tilde y}k=-\phi_1\,d_y .
\end{equation*}
For the Hessian, the product rule applied to $-\phi_1d_y$ gives
\begin{align*}
  \nabla^2_{\tilde y}k
  &=-\,d_y\bigl(\nabla_{\tilde y}\phi_1\bigr)^\top-\phi_1\,\nabla_{\tilde y}d_y\\
  &=-\,d_y\bigl(-\phi_2\,d_y\bigr)^\top-\phi_1\bigl(-D_\eta\bigr)\\
  &=\phi_2\,d_yd_y^\top+\phi_1D_\eta .
\end{align*}
With respect to $x$ the signs of the first derivatives change and the Hessian
is the same:
\begin{equation*}
  \partial_tk=\phi_1\,d_t,\qquad
  \nabla_yk=\phi_1\,d_y,\qquad
  \nabla^2_yk=\phi_2\,d_yd_y^\top+\phi_1D_\eta .
\end{equation*}

\subsection{Single application}

Inserting the derivatives in $\tilde x$ into
$\mathcal L_{\tilde x}k=\partial_{\tilde t}k-\tilde rk+\tilde\mu^\top\nabla_{\tilde y}k
+\tfrac12\operatorname{tr}(\tilde A\,\nabla^2_{\tilde y}k)$,
\begin{align*}
  \mathcal L_{\tilde x}k
  &=-\phi_1d_t-\tilde rk-\phi_1\,\tilde\mu^\top d_y
    +\tfrac12\phi_2\operatorname{tr}\bigl(\tilde Ad_yd_y^\top\bigr)
    +\tfrac12\phi_1\operatorname{tr}\bigl(\tilde AD_\eta\bigr).
\end{align*}
Using $\operatorname{tr}(\tilde Ad_yd_y^\top)=d_y^\top\tilde Ad_y$ and grouping,
\begin{equation*}
  F(x,\tilde x):=\mathcal L_{\tilde x}k
  =-\tilde r\,k-\phi_1\,\theta+\tfrac12\phi_2\,\Omega+\tfrac12\phi_1\,\tilde\tau_A,
\end{equation*}
with the three scalars
\begin{equation*}
  \theta=d_t+\tilde\mu^\top d_y,\qquad
  \Omega=d_y^\top\tilde A\,d_y,\qquad
  \tilde\tau_A=\operatorname{tr}\bigl(\tilde AD_\eta\bigr).
\end{equation*}

\subsection{Gradient of $F$}\label{app:gradF}

We now apply $\mathcal L_x$ to $F$, which requires $\partial_tF$, $\nabla_yF$
and $\nabla^2_yF$. In $F$, the quantities $\tilde r$, $\tilde\mu$, $\tilde A$
and $\tilde\tau_A$ depend only on $\tilde x$; the scalars $\theta$ and $\Omega$
depend on $x$ through $d_t$ and $d_y$:
\begin{align*}
  \nabla_y\theta&=D_\eta\tilde\mu, & \nabla_y\Omega&=2D_\eta\tilde A\,d_y, &
  \partial_t\theta&=\frac1{\eta_t^2}, & \partial_t\Omega&=0 .
\end{align*}
The product rule, term by term, gives
\begin{align*}
  \nabla_yF
  &=-\tilde r\bigl[\nabla_yk\bigr]
    -\bigl[\nabla_y\phi_1\bigr]\theta-\phi_1\bigl[\nabla_y\theta\bigr]\\
  &\quad+\tfrac12\bigl[\nabla_y\phi_2\bigr]\Omega
    +\tfrac12\phi_2\bigl[\nabla_y\Omega\bigr]
    +\tfrac12\tilde\tau_A\bigl[\nabla_y\phi_1\bigr]\\
  &=-\tilde r\bigl[\phi_1d_y\bigr]
    -\bigl[\phi_2d_y\bigr]\theta-\phi_1\bigl[D_\eta\tilde\mu\bigr]\\
  &\quad+\tfrac12\bigl[\phi_3d_y\bigr]\Omega
    +\tfrac12\phi_2\bigl[2D_\eta\tilde Ad_y\bigr]
    +\tfrac12\tilde\tau_A\bigl[\phi_2d_y\bigr].
\end{align*}
The terms proportional to $d_y$ are collected in a scalar $F_1$:
\begin{equation*}
  \nabla_yF=F_1\,d_y-\phi_1D_\eta\tilde\mu+\phi_2D_\eta\tilde A\,d_y,
\end{equation*}
\begin{equation*}
  F_1=-\tilde r\,\phi_1-\phi_2\,\theta+\tfrac12\phi_3\,\Omega+\tfrac12\phi_2\,\tilde\tau_A .
\end{equation*}
Note that $F_1$ is $F$ with every profile raised by one order
($k\to\phi_1$, $\phi_1\to\phi_2$, $\phi_2\to\phi_3$). The same computation applied
to $F_1$ raises the orders once more:
\begin{equation*}
  \nabla_yF_1=F_2\,d_y-\phi_2D_\eta\tilde\mu+\phi_3D_\eta\tilde A\,d_y,
\end{equation*}
\begin{equation*}
  F_2=-\tilde r\,\phi_2-\phi_3\,\theta+\tfrac12\phi_4\,\Omega+\tfrac12\phi_3\,\tilde\tau_A .
\end{equation*}
For the time derivative, with $\partial_t\phi_n=\phi_{n+1}d_t$,
\begin{align*}
  \partial_tF
  &=-\tilde r\bigl[\phi_1d_t\bigr]-\bigl[\phi_2d_t\bigr]\theta-\frac{\phi_1}{\eta_t^2}
    +\tfrac12\bigl[\phi_3d_t\bigr]\Omega+\tfrac12\tilde\tau_A\bigl[\phi_2d_t\bigr]\\
  &=F_1\,d_t-\frac{\phi_1}{\eta_t^2}.
\end{align*}

\subsection{Hessian of $F$}\label{app:hessF}

We differentiate the three terms of $\nabla_yF$ with respect to $y^\top$. For the
first, the product rule and $\nabla_yd_y=D_\eta$ give
\begin{align*}
  \frac{\partial}{\partial y^\top}\bigl(F_1\,d_y\bigr)
  &=F_1\,D_\eta+d_y\bigl(\nabla_yF_1\bigr)^\top\\
  &=F_1\,D_\eta+F_2\,d_yd_y^\top-\phi_2\,d_y\tilde\mu^\top D_\eta
    +\phi_3\,d_yd_y^\top\tilde AD_\eta .
\end{align*}
For the second and the third,
\begin{align*}
  \frac{\partial}{\partial y^\top}\bigl(-\phi_1D_\eta\tilde\mu\bigr)
  &=-D_\eta\tilde\mu\bigl(\nabla_y\phi_1\bigr)^\top
   =-\phi_2\,D_\eta\tilde\mu\,d_y^\top,\\[2pt]
  \frac{\partial}{\partial y^\top}\bigl(\phi_2D_\eta\tilde Ad_y\bigr)
  &=\phi_2\,D_\eta\tilde AD_\eta+D_\eta\tilde Ad_y\bigl(\nabla_y\phi_2\bigr)^\top\\
  &=\phi_2\,D_\eta\tilde AD_\eta+\phi_3\,D_\eta\tilde Ad_yd_y^\top .
\end{align*}
Adding the three contributions,
\begin{align*}
  \nabla^2_yF
  &=F_2\,d_yd_y^\top+F_1\,D_\eta+\phi_2\,D_\eta\tilde AD_\eta\\
  &\quad-\phi_2\bigl[D_\eta\tilde\mu\,d_y^\top+d_y\tilde\mu^\top D_\eta\bigr]
   +\phi_3\bigl[D_\eta\tilde Ad_yd_y^\top+d_yd_y^\top\tilde AD_\eta\bigr],
\end{align*}
which is symmetric, as it should be.

\subsection{Double application}

The three ingredients are
\begin{align*}
  \partial_tF&=F_1\,d_t-\frac{\phi_1}{\eta_t^2},\\
  \nabla_yF&=F_1\,d_y-\phi_1D_\eta\tilde\mu+\phi_2D_\eta\tilde A\,d_y,\\
  \nabla^2_yF&=\text{as above}.
\end{align*}
The drift term is
\begin{equation*}
  \mu^\top\nabla_yF=F_1\,\mu^\top d_y-\phi_1\,\mu^\top D_\eta\tilde\mu
  +\phi_2\,\mu^\top D_\eta\tilde Ad_y .
\end{equation*}
For the diffusion term we use $\operatorname{tr}(A\,a\,b^\top)=b^\top A\,a$ and
the symmetry of $A$, $\tilde A$ and $D_\eta$; with $\tau_A=\operatorname{tr}(AD_\eta)$,
\begin{align*}
  \tfrac12\operatorname{tr}\bigl(A\,\nabla^2_yF\bigr)
  &=\tfrac12F_2\,d_y^\top Ad_y+\tfrac12F_1\,\tau_A
    +\tfrac12\phi_2\operatorname{tr}\bigl(AD_\eta\tilde AD_\eta\bigr)\\
  &\quad-\phi_2\,d_y^\top AD_\eta\tilde\mu
    +\phi_3\,d_y^\top AD_\eta\tilde Ad_y .
\end{align*}
Summing $\partial_tF-rF+\mu^\top\nabla_yF+\tfrac12\operatorname{tr}(A\,\nabla^2_yF)$
and grouping the terms in $F_1$,
\begin{align*}
  \mathcal L_x\mathcal L_{\tilde x}k
  &=-r\,F+F_1\Bigl(d_t+\mu^\top d_y+\tfrac12\tau_A\Bigr)
    +\tfrac12F_2\,d_y^\top Ad_y-\frac{\phi_1}{\eta_t^2}\\
  &\quad-\phi_1\,\mu^\top D_\eta\tilde\mu
    +\phi_2\,\mu^\top D_\eta\tilde Ad_y
    -\phi_2\,d_y^\top AD_\eta\tilde\mu\\
  &\quad+\phi_3\,d_y^\top AD_\eta\tilde Ad_y
    +\tfrac12\phi_2\operatorname{tr}\bigl(AD_\eta\tilde AD_\eta\bigr).
\end{align*}
This is the result stated in the double-application proposition of the main text; it involves the kernel
only through $\phi_1,\dots,\phi_4$.

\subsection{Coincident points}\label{app:diag}

On the diagonal of the Gram matrix, $x=\tilde x$: then $\rho=0$, $d_t=0$,
$d_y=0$, hence $\theta=\Omega=0$, and $\mu=\tilde\mu$, $A=\tilde A$. If the
profiles $\phi_3$ and $\phi_4$ are bounded, every term containing $d_t$ or $d_y$
simply vanishes. If they are singular, as for the Matérn kernel, the products
still vanish provided $\phi_3=O(\rho^{-1})$ and $\phi_4=O(\rho^{-3})$, because
$d_t$, $d_y$ and $\theta$ are $O(\rho)$ and $\Omega$ is $O(\rho^2)$. In both cases
\begin{align*}
  \mathcal L_x\mathcal L_{\tilde x}k\Big|_{x=\tilde x}
  &=-r\,F^{\circ}+\tfrac12F_1^{\circ}\,\tau_A-\frac{\phi_1(0)}{\eta_t^2}
    -\phi_1(0)\,\mu^\top D_\eta\mu\\
  &\quad+\tfrac12\phi_2(0)\operatorname{tr}\bigl(AD_\eta AD_\eta\bigr),
\end{align*}
with $F^{\circ}=-r\,k(0)+\tfrac12\phi_1(0)\tau_A$ and
$F_1^{\circ}=-r\,\phi_1(0)+\tfrac12\phi_2(0)\tau_A$. In an implementation this limit
must be imposed explicitly on the diagonal when the profiles are singular.

\subsection{Profiles of the Matérn 5/2 kernel}

For $k(\rho)=a^2\bigl(1+\sqrt5\rho+\tfrac53\rho^2\bigr)e^{-\sqrt5\rho}$,
\begin{align*}
  k'(\rho)&=a^2\Bigl(\sqrt5+\tfrac{10}{3}\rho\Bigr)e^{-\sqrt5\rho}
           -a^2\sqrt5\Bigl(1+\sqrt5\rho+\tfrac53\rho^2\Bigr)e^{-\sqrt5\rho}\\
          &=-\tfrac53a^2\,\rho\,(1+\sqrt5\rho)\,e^{-\sqrt5\rho},
\end{align*}
so that
\begin{equation*}
  \phi_1=-\tfrac53a^2(1+\sqrt5\rho)e^{-\sqrt5\rho}.
\end{equation*}
Differentiating again,
\begin{align*}
  \phi_1'&=-\tfrac53a^2\Bigl[\sqrt5-\sqrt5(1+\sqrt5\rho)\Bigr]e^{-\sqrt5\rho}
          =\tfrac{25}{3}a^2\,\rho\,e^{-\sqrt5\rho},
  &\phi_2&=\tfrac{25}{3}a^2e^{-\sqrt5\rho},\\
  \phi_2'&=-\tfrac{25\sqrt5}{3}a^2e^{-\sqrt5\rho},
  &\phi_3&=-\tfrac{25\sqrt5}{3}a^2\,\frac{e^{-\sqrt5\rho}}{\rho},
\end{align*}
and
\begin{equation*}
  \phi_3'=\tfrac{25\sqrt5}{3}a^2\Bigl(\frac{\sqrt5}{\rho}+\frac1{\rho^2}\Bigr)e^{-\sqrt5\rho},
  \qquad
  \phi_4=\tfrac{25}{3}a^2\Bigl(\frac5{\rho^2}+\frac{\sqrt5}{\rho^3}\Bigr)e^{-\sqrt5\rho}.
\end{equation*}
The profiles $\phi_1$ and $\phi_2$ are bounded, with
$\phi_1(0)=-\tfrac53a^2$ and $\phi_2(0)=\tfrac{25}{3}a^2$; $\phi_3$ and $\phi_4$ are
singular at the origin with the rates $\rho^{-1}$ and $\rho^{-3}$ required above.

\subsection{Profiles of the Gaussian kernel}

For $k(\rho)=a^2e^{-\rho^2/2}$ one has $k'(\rho)=-\rho\,k(\rho)$, so
$\phi_1=-k$. If $\phi_n=(-1)^nk$, then $\phi_n'=(-1)^{n+1}\rho\,k$ and
$\phi_{n+1}=(-1)^{n+1}k$. Hence
\begin{equation*}
  \phi_1=-k,\qquad\phi_2=k,\qquad\phi_3=-k,\qquad\phi_4=k .
\end{equation*}
All profiles are bounded, with $\phi_n(0)=(-1)^na^2$, so the general formulas
can be evaluated everywhere, including on the diagonal.